\documentstyle[12pt]{article}

\advance\hoffset by -1.1truecm
\advance\voffset by -1.0truecm

\def\be{\begin{equation}}
\def\ee{\end{equation}}
\def\ba{\begin{eqnarray}}
\def\ea{\end{eqnarray}}
\def\v#1{\vert #1 \rangle}
\def\sp#1#2{\langle#1\vert#2\rangle}

\def\me#1#2#3{\langle #1 \vert #2 \vert #3 \rangle}

\newcommand{\x}{{\bf x}}
\newcommand{\p}{{\bf p}}
\newcommand{\etab}{{\bar{\eta}}}
\newcommand{\detab}{{\partial_{\etab}}}
\newcommand{\deta}{{\partial_{\eta}}}
\newcommand{\xl}{{\frac{x}{L}}}
\newcommand{\sn}{\smallskip\newline}
\newcommand{\mn}{\medskip\newline}
\newcommand{\bn}{\bigskip\newline}
\newcommand{\mbo}{{\mbox{ }}}
\newcommand{\qqq}{ \qquad \qquad \qquad }
\def\detapb#1{{\partial_{\etab^{#1}}}}
\def\detapp#1{{\partial_{\eta^{#1}}}}

\begin{document}

\title{Quantum Field Theory with Nonzero Minimal Uncertainties
in Positions and Momenta}
\author{Achim Kempf\thanks{supported by Studienstiftung des Deutschen
Volkes, BASF-fellow}\\
Department of Applied Mathematics \& Theoretical Physics\\
University of Cambridge, Cambridge CB3 9EW, U.K.\\
 {\small Email: a.kempf@damtp.cambridge.ac.uk}}

\date{}
\maketitle
\vskip-9cm
\hskip11.2cm
{{\bf DAMTP/94-33 \rm}}
\vskip0.31cm
\hskip11.2cm hep-th/9405067
\vskip8.6cm

\begin{abstract}
A noncommutative geometric generalisation
of the quantum field theoretical framework is developed by
generalising the Heisenberg commutation relations.
There appear nonzero minimal uncertainties in positions and in
momenta. As the main result it is shown with the example of
a quadratically ultraviolet divergent
graph in $\phi^4$ theory that
nonzero minimal uncertainties in positions
do have the power to regularise.
These studies are motivated with the ansatz that nonzero
minimal uncertainties in
positions and in momenta arise from gravity.
Algebraic techniques are used that have been developed in the
field of quantum groups.
\end{abstract}
\newpage
\tableofcontents
\newpage

\section{Introduction}
Over the past few years there has been
considerable progress in the mathematics
of noncommutative geometry, and in particular in the field
of braided geometry and
quantum groups. It has been suggested that the introduction of
new 'noncommutative geometric' methods
into quantum theory could
lead to a deeper understanding
of spacetime geometry and thus even to a new connection
between quantum theory and gravity.

In this paper we study in fully explicit examples how the quantum field
theoretical framework can be generalised for
noncommutative position and momentum spaces.
Instead of generalising by modifying the linear superposition
principle, see e.g. \cite{w}, our programme here is
to work with generalisations of the Heisenberg commutation relations.
It is shown that there then naturally appear nonzero minimal
uncertainties, separately in position and in momentum measurements.
 We demonstrate that the appearance of e.g.
a nonzero minimal uncertainty in position has the power to regularise
ultraviolet divergencies.

The results of the present paper can thus
be viewed as being an approach
towards a new regularisation method. This should in itself be interesting.
However, studying the effects of nonzero minimal position- or
momentum uncertainties in quantum field theory,
we do have in mind the idea that gravity could
be introduced as the cause of these uncertainties.
Let us explain this motivation.
\mn
The quantum field theoretical framework, as it is used e.g. for the
standard model is certainly very successful
from a macroscopic scale down to at least the scale of today's high energy
experiments. However, it is clear that the ordinary
quantum field theoretical framework
cannot describe the physics of the large scales and also not the
physics of the very small scales (like the Planck length of $10^{-35}$m),
since in both cases the spacetime structure becomes nontrivial:

In large distances the curvature
of spacetime becomes significant and there is
on a general curved spacetime
no notion of a plane wave.
This means that there is a limit
to the precision to which momenta can be described.
We will make the ansatz that this can be expressed as a (dynamical)
nonzero minimal uncertainty in momentum measurements, where we apply the
usual definition of the uncertainty in an observable:
$$(\Delta p)^2_{\v{\psi}} = \langle \psi \vert ( \p - \langle \psi \vert
\p \vert \psi \rangle )^2\vert \psi \rangle $$
On the other hand, as one tries to measure
very small distancies, one needs particles of very high energies,
which should eventually also lead to perturbing gravity effects.
Therefore we expect there also to be a limit to
the precision to which positions can be described.
We will make the ansatz that this can be
expressed as a (dynamical) nonzero minimal uncertainty
in position measurements where we will again apply the
usual definition for the uncertainty in an observable.

Although it is quite obvious that in nature positions and momenta can
not be measured arbitrarily precisely, even in principle,
it is however a
nontrivial ansatz to assume that this can be expressed as
nonzero minimal uncertainties, applying the ordinary quantum
theoretical
definition of uncertainties in observables.
\bn
The ordinary quantum field theoretical framework
does not imply a limit to the precision
to which positions or momenta could be measured. Mathematically, the
ordinary framework deals with
arbitrarily small and arbitrarily large scales in position and in
momentum space. However the framework does not describe
gravity effects that appear
in nature at the large and at the
very small scales. Instead, the framework allows for infrared
 and ultraviolet
divergencies to occur at the extreme scales.

It appears that the ordinary quantum field theoretical framework
suffers from problems, the cause of which can be
traced back to quantum mechanics: The (associative)
Heisenberg algebra, generated by the hermitean
position and momentum operators $\x$ and $\p$, obeying $[\x,\p]=i\hbar$,
is represented on the vector space of states. The observables
$\x$ and $\p$ have eigenstates, which can be written as
$\delta$-functions and plane waves.
Now in general, for any eigenstate $\v{a}$ of an observable $A$ with
$A.\v{a}= a \v{a}$, the uncertainty in the measurement of
$A$ of course vanishes:
$$(\Delta A)^2_{\v{a}} = \langle a\vert (A-\langle a,A.a\rangle )^2\vert
a\rangle = 0 $$
The existence of $\x$- or $\p$- eigenstates
in the representation of the Heisenberg algebra thus implies that the
usual quantum theoretical framework allows in principle for
the description of arbitrarily precisely measured positions
or momenta.

This does not cause many problems on the level of quantum mechanics,
apart from the fact that the $\x$- or $\p$- eigenstates strictly speaking
are not eigenvectors. Dirac $\delta$- 'functions' and plane waves are
not square integrable and thus not contained in the Hilbert space.
Difficulties with the multiplication
of $\delta$-functions are related to that.

The problem is, that the quantum
field theoretical framework is built on position space, or
equivalently, on momentum space. The notion of points in position or
in momentum space, i.e. of particles in position or momentum eigenstates,
is thus implemented from the very beginning. Independent of the actual
interactions, it is thus the framework itself that pretends that position or
momentum measurements of arbitrary precision are possible.
\mn
We therefore suggest as a possible cure for this problem
to generalise the quantum field theoretical framework such, that
it no longer allows for arbitrarily precise measurements of
positions or momenta. To this end
the quantum fields first have to be redefined on a space
alternative to position or momentum space.
It is essential to work in a basis other
than the position or the momentum eigenbasis
because only then can we abandon the
very existence of $\x$- or $\p$- eigenstates, i.e. the
existence of 'points' in position or momentum space.

In a generalised Heisenberg algebra not only the positions and momenta
do not commute, now the positions and
the momenta may no longer commute even
among themselves. While quantisation abandoned the
classical phase space, this also abandons the configuration
and the momentum space.

The generalisation parameters $q_i$ that control the
commutation relations have one special value, say $q_i=1$,
that corresponds to the ordinary quantum field theoretical framework.
The generalised Heisenberg algebra, then involving a length scale,
should imply nonzero minimal uncertainties
in positions and momenta.
Intuitively a nonzero minimal
uncertainty e.g. in positions is of course like a short-distance cutoff.
It is however nontrivial to show that a nonzero minimal
$x$- uncertainty really leads to the
regularisation of otherwise ultraviolet divergent graphs, with
the divergency only to be recovered when
the generalisation parameters approach $q_i=1$.

The finite values of the loop integrations together with the
sizes of the minimal uncertainties depend
on the generalisation parameters.
As we said, ultimately we expect nonzero
minimal uncertainties in positions and momenta to be
arising from gravity. Thus, the further idea is
to try to relate the generalisation
parameters (which then have to be dynamical)
to gravitational degrees of freedom.
In this way gravity effects might actually regularise
quantum field theory. To make contact with gravity is beyond the scope of
the present paper. We will only sketch a possible ansatz in Section 4.2.
\bigskip\newline
The programme for our studies is as follows:
\sn
First we reformulate the ordinary quantum field theoretical framework
such that the quantum fields no longer live on position or momentum
space. Instead of the basis of $\delta$-functions or plane
 waves we will here choose
to define them on a basis of Bargmann Fock functions.
Other choices are possible, the main objective is,
as we said, to achieve independence
of the $x$- eigenbasis and the $p$- eigenbasis.
Up to this point we only reformulate the ordinary formalism, with
the fields now living on an alternative space such as Bargmann Fock
space, rather than on
position or momentum space.
The unitary transformations to and from quantum fields
that are defined on
position or momentum space are given.
\sn
Secondly then, we generalise the Heisenberg algebra.
In principle we are interested in generalising its
commutation relations as far as consistently possible.
For explicit calculations Hilbert space representations
for the generalised Heisenberg algebras are needed.
We postpone the fully general construction
of Hilbert space representations.
As we will see, large classes of generalised Heisenberg commutation
relations can be represented on certain generalised
Bargmann Fock spaces. We will here consider only these restricted
generalisations of the Heisenberg
algebras, since for these we can easily find Hilbert space
 representations, namely by using quantum group methods.
\mn
This strategy of first reformulating the ordinary quantum
theoretical formalism in an alternative (e.g. the
Bargmann Fock-) basis and then secondly generalising the Heisenberg
algebra, we will be
following in Section 2 for quantum mechanics and in section 3 for
quantum field theory.
\sn
The set of generalised commutation relations that we can
deal with up to now is, as we will see in Section 4,
not yet large enough for an attempt to describe gravity.
These restricted generalisations do however
already show features that we expect to be arising
from gravity:
With the generalised commutation relations, the position
 and momentum operators
do no longer have eigenfunctions in the representations
of the Heisenberg algebra. Instead,
there are now nonzero minimal uncertainties in $\x$- and in $\p$-
measurements.
Intuitively one can expect that this
implies short distance cutoffs in position and in momentum space.
Indeed, working with $\phi^4$ theory
we will show for the example of a quadratic ultraviolet divergency
that the introduction of nonzero
minimal uncertainties in positions does indeed have the power
 to regularise.

The significance of this
result is, that it supports the idea that gravity effects can
be introduced into quantum field theory in such a way that they
actually regularise quantum field theory.
\bigskip\newline
As we said, we will here work only with specific examples of a
general approach. With further Hilbert space
representations of more generalised Heisenberg algebras
it should be possible to straightforwardly
follow the same programme to
formulate the correspondingly generalised quantum theoretical
framework.
\sn
The representations that we will use here for our examples of
generalised Heisenberg commutation relations are
representations on so called $q$- Bargmann Fock spaces,
 which have been developed in the context of quantum groups.
We do therefore naturally include the case when
the commutation relations are preserved by a
quantum group of linear quantum canonical transformations.
However, let us stress that we are covering
a more general case and
in particular that we do not have a quantum group of external
symmetry. The wave functions will be commutative and we will
not be dealing with particles that obey braid statistics.
We are using the algebraic techniques that have been developed
in the field of quantum groups merely as a powerful tool for the
construction of certain Hilbert space representations.
\mn
Standard references for the mathematics of quantum groups are
\cite{sweedler}-\cite{dc-wess}.
The $SU_q(n)$- symmetric Bargmann Fock spaces that we will
be using have been constructed in
\cite{ak-lmp-bf,ak-jmp-bf}, for questions of uniqueness see also
\cite{ak-proc-ny,ak-proc-berlin}. Using this formalism for
nonrelativistic quantum mechanics the study of
nonzero minimal uncertainties in $\x$ and $\p$ has been worked out
in \cite{ak-jmp-ucr}.
\bn
There are many other approaches towards the implementation
of several kinds of
noncommutative geometry into quantum theory or gravity, see e.g. [13-30].
In particular, some of the quantum group symmetric
commutation relations that we use here, which
were derived in the $R$-matrix approach in \cite{ak-lmp-bf}, had already,
 as I learned later, appeared in a different approach in \cite{woro}.
However, our generalised Heisenberg algebras with their
Bargmann Fock representations,
the derivation of the nonzero minimal
uncertainties, and our field theoretical studies here, are new.
An interesting new alternative approach is \cite{doplicher 94/6}.
\mn
The idea of the existence of a minimal observable length is of course
not new, see e.g. \cite{townsend,jaeckel}, and for a
recent survey \cite{garay}.
Interestingly an effective uncertainty
relation implying nonzero minimal uncertainties in positions has
been derived in the context of string theory, where it
effectively arises from the string size, see e.g.
\cite{amati,konishi,maggiore} and references therein.
That uncertainty relation is a special case of
the uncertainty relation
of our framework. This may indicate that our framework could serve as
an effective
theory of an underlying theory of explicitly non-pointlike particles,
such as string theory.
The connection between the generalisation parameters and the gravitational
degrees of freedom would then of course be quite indirect.

\section{Quantum mechanics with nonzero minimal uncertainties}
In the two sections 2.1 and 2.2 of this chapter we study quantum
mechanics and our generalisations of it in a parallel way.

We choose to work in the Bargmann Fock basis
of the (Hilbert) space of states. In order to write operators as
integral kernels we use
a new formula for the integral over two Bargmann Fock functions.
This algebraic integral also works in the generalised case.
Studying the position and momentum operators we give the
unitary transformations that map from the Bargmann Fock basis to the
position or the momentum eigenbasis and vice versa. In the generalised case
there are no more position or momentum eigenstates and the uncertainty
relation implies nonzero minimal uncertainties in positions and momenta.
We develop new algebraic techniques which will allow
us to generalise the
notion of pointwise multiplication to the noncommutative situation.
This will be crucial in section 3 for the description
of local interactions in quantum field theory.

\subsection{Quantum mechanics in the Bargmann Fock basis}
In the following four subsections we deal with ordinary quantum
mechanics. After a brief review of the notion of Bargmann
Fock representation we
develop some useful new formulas for the transformations from
Bargmann Fock space to position and momentum space,
and vice versa. We then apply them to
the translation of the
notion of pointwise multiplication from position space into
Bargmann Fock space.

\subsubsection{Bargmann Fock space}
The Heisenberg algebra, generated by the observables $\x$ and $\p$,
obeying $[\x,\p]=i\hbar$, is represented on the space
of states in which one usually chooses a basis of position or momentum
eigenvectors $\v{x}$, $\v{p}$. The states are then written expanded as
wave functions $\psi_x(x):= \sp{x}{\psi}$ or $\psi_p(p):= \sp{p}{\psi}$
and $\x,\p$ act on them as multiplication and differentiation operators
$x, -i\hbar\partial_x$ or $i\hbar\partial_p, p$.
The coordinates $\sp{x}{\psi}$ of a state $\v{\psi}$ in the
basis of position eigenvectors are unitarily transformed
into its coordinates $\sp{p}{\psi}$
in the basis of momentum eigenvectors using
$\sp{p}{x}=(2\pi \hbar)^{-1/2} e^{-ipx/\hbar}$, i.e.
by Fourier transformation
$
\sp{p}{\psi} = \int_{-\infty}^{+\infty} dx \sp{p}{x}\sp{x}{\psi}
$.

It is well known that alternatively
 one can e.g. choose the basis of eigenvectors of the
operator $\etab \detab$ where:
\be
\etab := \frac{1}{2L} \x  -\frac{iL}{\hbar} \p \mbox{ \quad and \quad }
\detab := \frac{1}{2L} \x  +\frac{iL}{\hbar} \p
\label{etadef}
\ee
Here, with $L$ we introduce an arbitrary positive length.
One checks that
$\etab$ and $\detab$ indeed obey
$\detab \etab -\etab \detab = 1$, which is of the form of
a Leibniz rule. For the Bargmann Fock operator we
readily write down the countable set of eigenvectors
$ \etab \detab \v{\etab^{n}} = n \v{\etab^{n}}$
with $ n=0,1,2,... $
which are found to have the scalar product:
\be
\sp{\etab^n}{\etab^m} = n! \mbox{ }\delta_{n,m}
\label{sp}
\ee
With the definitions
$
\v{a \etab^n + b \etab^m} := \v{a \etab^n} + \v{b \etab^m} $ and $
 a \v{\etab^n} := \v{a \etab^n}
$,
arbitrary states $\v{\psi}$ can be written as polynomials
or power series $$\v{\psi(\etab)} =
\v{\sum_{r=0}^{\infty}\psi_r\frac{\etab^r}{\sqrt{r!}}}$$
The action of the generators $\x$ and $\p$ of the
Heisenberg algebra is then obtained from Eqs.\ref{etadef}, namely
\be
\x = L(\etab + \detab)
\qquad \qquad
\p = \frac{i\hbar}{2L}(\etab - \detab)
\label{xpdef}
\ee
so that e.g.:
\be
\x \v{\psi(\etab)} = L \v{\etab \psi(\etab) + \detab \psi(\etab)}
\ee
We have
\be
\sp{x}{\etab^n} = \sqrt{n!} (2\pi L^2)^{-1/4} (x/2L - L\partial_x)^n
e^{-\frac{1}{4}\left(\frac{x}{L}\right)^2}
\label{unha1}
\ee
which are, up to a factor, the Hermite functions.
The coordinates $\psi_n=\sp{\etab^n}{\psi}/\sqrt{n!}$
of a function $\v{\psi}$, in the orthonormalised
basis of monomials $\etab^n/{\sqrt{n!}}$,
are unitarily transformed into its coordinates
$\psi_x(x)=\sp{x}{\psi}$ in the basis of position eigenvectors:
\be
\sp{x}{\psi} = \sum_{n=0}^{\infty} \sp{x}{\etab^n}
 \sp{\etab^n}{\psi} (n!)^{-1}
\label{unha2}
\ee
To summarise, we are presently considering three different bases of our
space of states. In particular, the set of states $\v{\psi}$
which yield a square integrable wave function
$\psi_x(x) = \sp{x}{\psi}$
(and thus also a square integrable
$\psi_p(p) = \sp{p}{\psi}$) forms a Hilbert space $\cal{H}$ and these states
can also be written as
holomorphic, weighted square integrable
'Bargmann Fock functions' $\v{\psi(\etab)}$ and vice versa.
Recall that the scalar
product of two arbitrary states, when given as
Bargmann Fock functions, can be calculated e.g. as
\be
\sp{\psi(\etab)}{\phi(\etab)} = \frac{1}{2\pi i} \int d\eta d\etab\mbo
\overline{\psi(\etab)}\mbo  e^{-\etab \eta}\mbo \phi(\etab)
\label{spalt}
\ee
where the $\psi(\etab)$ and $\phi(\etab)$ on the rhs are now to be read
as polynomials or power series in ordinary complex variables rather
than in Bargmann Fock operators $\etab$.
\mn
We should here clarify an important point:

A 'wave function' gives
the coefficients of a state in a particular basis. E.g.
$\psi(x) = \sp{x}{\psi}$ and $\psi(p) = \sp{p}{\psi}$ are the coefficients
of the state $\vert \psi \rangle $ in the position
or momentum eigenbasis, i.e. they
are the position or momentum wave functions. However, Bargmann Fock
functions $\psi(\etab)$ are
{\bf not\rm}\footnote{$ \psi(\etab) $ is \bf not \rm the coefficient
of the state $\vert \psi \rangle$ in a '$\etab$-eigenbasis'. An
$\etab$-eigenbasis does not exist.
We are also not dealing with the (overcomplete)
coherent states, which would be eigenstates of the operator
$\detab$.} wave functions! $\psi(\etab)$
is a function, or more precisely, it is a polynomial
or power series in the Bargmann Fock operator
$\etab$ and thus it is itself an operator on the representation space.
Applied on the normalised state $\vert 1\rangle$ (defined through
$\detab. \vert 1\rangle = 0$) one obtains a state $\psi(\etab) \vert
1\rangle$ or $\vert \psi(\etab) \rangle$. This state is usually
also simply denoted by $\psi(\etab)$. (Thus, in particular
the state $\v{\etab}$ is defined as $\etab.\v{1}$ and it is not
an eigenstate to the operator $\etab$.)

On the other hand, the Bargmann Fock operator $\etab\detab$ is self-adjoint
with the eigenbasis $ \{\v{(r!)^{-1/2} \etab^r} \}$ $(r=0,1,2,...)$.
Just like the position or momentum wavefunctions $\sp{x}{\psi}$
or $\sp{p}{\psi}$ that we get
as the coefficients of a state in the $\x$ or $\p$- eigenbases
$\{ \v{x}\}$ or $\{\v{p}\}$, we can therefor now also calculate a
Bargmann Fock \it wave function\rm:
\be
\psi_r = \sp{(r!)^{-1/2} \etab^r}{\psi}\qquad \qquad r=0,1,2,...
\ee
Unlike the $\x$ or the $\p$ eigenbases, the eigenbasis of the Bargmann Fock
operator $\etab\detab$ is countable, so that our
Bargmann Fock \it wave functions \rm are actually sequences rather than
functions (we are making use of the fact that the
Hilbert space of states is
separable).

\label{cadi}
Let us carefully distinguish the Bargmann Fock \it functions \rm
$\psi(\etab)$ or $\v{\psi(\etab)}$, which denote operators on
or vectors in the representation space, from the
Bargmann Fock \it wave functions \rm $\psi_r$,
which are the complex number valued
coefficients of a state, when expanded in the eigenbasis of the
operator $\etab\detab$.

This is to be kept in mind when we will be speaking, like in the
literature, about Bargmann Fock
functions $\psi(\etab)$ on 'Bargmann Fock space'.
It should now be clear that what is meant is not a wave function,
but a state, namely the state $\psi(\etab).\v{1}$.

In our generalised framework we will be dealing with Bargmann Fock
\it functions \rm $\psi(\etab_i)$ where e.g.
the Bargmann Fock operators $\etab_i$ no longer commute. Nevertheless,
the Bargmann Fock \it wave functions \rm $\psi_r$,
i.e. the coefficients of the
states expanded in the eigenbasis of $\etab_i\detab_i$ remain of
course complex number valued, and thus the Bargmann Fock \it wave
functions \rm remain commutative. Further then this will mean that
when we generalise e.g. bosonic $\phi^4$- theory, the fields that
we sum over in the path integral are strictly commutative. We
will see that even in the case when
the commutation relations of the
Heisenberg algebra are conserved under the action of a
quantum group, we will not be dealing with particles that
obey braid statistics\footnote{Recall that the 'universal R-matrix'
that determines the quasitriangular structure of a quantum group
defines a set of intertwining morphisms of quantum group modules,
which forms itself a representation of the braid group.}.

\subsubsection{Integral kernels}
We recall that, using the weighted complex integral Eq.\ref{spalt},
operators can also be written as integral kernels: Let $P(\etab,\detab)$
be an operator on the Hilbert space, expressed in
terms of the Bargmann Fock operators $\etab$ and $\detab$. (Using
Eqs.\ref{xpdef} it could of course also be written in terms of the
position and momentum operators.)
\mn
The integral kernel $G_P(\etab^{\prime},\eta)$ of $P$ is then a function
in the $\etab^{\prime}$ and $\eta$, which
lives in a primed and an unprimed
copy of the function space. Integrating $G_P$
over an arbitrary Bargmann Fock function
$\psi(\etab)$ yields the action of the operator $P$ on this
Bargmann Fock function:
\be
P(\etab^{\prime},\partial_{\etab^{\prime}}) \psi(\etab)
= \frac{1}{2\pi i}
\int d\etab d\eta\mbo G_P(\etab^{\prime},\eta)\mbo
e^{-\etab \eta} \mbo \psi(\etab)
\ee
The integral kernel of the identity operator is
$
G_1(\etab^{\prime},\eta) = e^{\etab^{\prime} \eta}
$
and e.g. the integral kernel of the time evolution operator
$
U = e^{-i(t_f-t_i)\omega \etab\detab}
$
of the harmonic oscillator, is the Greens function:
\be
G_U(\etab^{\prime},\eta) =
e^{\etab^{\prime}\eta e^{-i\omega (t_f-t_i)}}
\ee
\subsubsection{Positions and momenta}
The operators $\x$ and $\p$ have (non-normalisable) eigenvectors,
namely '$\delta$- functions' and plane waves, which are the 'points'
in position or momentum space. How do these states read
in the Bargmann Fock basis, and in general, how is a wave function
that is given on position space transformed into Bargmann Fock space?
For practical purposes the use of Eqs.\ref{unha1},\ref{unha2} for the
transformation from position space into Bargmann Fock space
 is rather inconvenient. Let us therefore, starting from known
expressions, develop some new and
more direct formulas for the transformations from the Bargmann
Fock basis to the position or momentum basis and vice versa.

Let a state $\v{\psi}$ be given in the position basis by its wave
function $\psi_x(x)=\sp{x}{\psi}$.
We can then calculate its Bargmann Fock
function $\v{\psi}=\v{\psi(\etab)}$ as follows:
\be
\psi(\etab) = (2\pi L^2)^{-1/4} \int_{-\infty}^{+\infty} dx\mbox{ }
e^{-\frac{1}{2}\etab^2+\etab\xl -\frac{1}{4}(\xl)^2} \psi_x(x)
\label{g1}
\ee
This result\footnote{Recall that this means that
$\psi(\etab)$, read as a function of
the Bargmann Fock operator $\etab$, applied on $\v{1}$ yealds the state
which has the position wave function $\psi_x(x)$.}
was obtained in \cite{bargmann},
see also \cite{perelomov} and
references therein.\newline
The inverse transformation is:
\be
\psi_x(x) = (8\pi^3L^2)^{-1/4}\int_{-i\infty}^{+i\infty} d\etab \mbox{  }
e^{\frac{1}{2}\etab^2-\etab\xl +\frac{1}{4}(\xl)^2} \psi(\etab)
\label{g2}
\ee
To see this, a short calculation shows, that
the Bargmann Fock function $\psi(\etab):=1$ is mapped
onto the position wave function
$$
\psi_x(x)=(2\pi L^2)^{-1/4}\mbo e^{-1/4\mbox{ }(x/L)^2}
$$
and vice versa.
The induction is then completed by showing that multiplying the
Bargmann Fock function with $\etab$ just amounts to the action
of $(x/L -2L\partial_x)/2$ on the wave function in position space.

These formulas, connecting the position space with the 'Bargmann Fock
space', are the analogue of the fourier transformation
formulas connecting
the position space with the momentum space.
Analogously one
derives similar formulas that connect the Bargmann Fock space directly
to momentum space:
\be
\psi(\etab) = \left(\frac{2L^2}{\pi \hbar^2}\right)^{1/4}
\int_{-\infty}^{+\infty} dp\mbo e^{\frac{1}{2}\etab^2 +
2i\etab\frac{Lp}{\hbar}-\left(\frac{Lp}{\hbar}\right)^2} \psi_p(p)
\label{mombf}
\ee
with the inverse:
\be
\psi_p(p) = \left(\frac{L^2}{2\pi^3 \hbar^2}\right)^{1/4}
\int_{-\infty}^{+\infty} d\etab\mbo e^{-\frac{1}{2}\etab^2 -
2i\etab\frac{Lp}{\hbar}+\left(\frac{Lp}{\hbar}\right)^2} \psi(\etab)
\ee
For the proof, note that $\sp{p}{1} =
\left(\frac{2L^2}{\pi \hbar^2}\right)^{1/4}
e^{-\left(\frac{Lp}{\hbar}\right)^2}$.
\sn
Let us remark that from Eqs.\ref{g1},\ref{g2} immediately follows
that the transformation
\be
\tilde{f}(y):=\int_{-\infty}^{+\infty} dx \mbox{ }
e^{-\frac{1}{L^2}(x-y)^2} f(x)
\label{uns}
\ee
which yields a 'gaussian-diluted' function has actually an inverse:
\be
f(x)=\frac{1}{\pi L^2} \int_{-\infty}^{+\infty} dy\mbox{ }
e^{+\frac{1}{L^2}(x-iy)^2} \tilde{f}(iy)
\label{revuns}
\ee
\sn
Using Eq.\ref{g1} we calculate the Bargmann Fock
function $\psi_{(x_0)}(\etab)$ of
e.g the $\x$-eigenvector with eigenvalue $x_0$, i.e. of
the $\delta$- function at $x_0$ on position space:
\be
\psi_{(x_0)}(\etab) =
(2\pi L^2)^{-1/4} e^{-\frac{\etab^2}{2} +
\etab \frac{x_0}{L} - \frac{1}{4}(\frac{x_0}{L})^2}
\label{bfdelta}
\ee
Now, putting this function into Eq.\ref{spalt} as one of its two
arguments, we calculate the scalar product of a state,
with a $\delta$- function in position space. This yields
another formula for the transformation from Bargmann Fock to
position space:
\be
\psi_x(x) = \frac{(2\pi L^2)^{-1/4}}{2\pi i} \int d\eta d\etab
e^{-\etab \eta -\frac{\eta^2}{2}+\eta \xl
-\frac{1}{4}(\xl)^2} \psi(\etab)
\label{rueck2}
\ee
In Bargmann Fock space the scalar product of states can actually also be
expressed purely algebraically, namely in terms of
derivatives \cite{ak-lmp-bf}:
\be
\sp{\psi}{\phi} = \overline{\psi(\etab)}\mbox{ }
e^{\partial_{\eta} \partial_{\etab}}\mbox{ }
\phi(\etab)\mbo \vert_{\eta=0=\etab}
\label{spneu}
\ee
Here the exponential is defined through its power series i.e.
$e^{\partial_{\eta} \partial_{\etab}} =
\sum_{r=0}^{\infty}
\frac{\partial_{\eta} \partial_{\etab}}{r!}$.
where the derivatives $\deta$ act from the right.
The evaluation procedure is to carry out the differentiations
and then to set $\eta$ and $\etab$ equal to zero. The remaining
number is the value of the scalar product.
This can be done
algebraically by using the Leibniz
rule $\partial_{\etab}\etab - \etab \partial_{\etab} = 1$ and
its complex conjugate $\eta \partial_{\eta} - \partial_{\eta} \eta =1$.
Note that we are using the left derivative $\detab$ and the right
derivative $\deta$, which means explicitly that e.g.
\begin{eqnarray*}
\detab \etab^2 &=& \detab \etab \etab = (\etab \detab +1)\etab
= \etab \detab \etab +\etab\\
 &=& \etab (\etab \detab +1) +\etab
=\etab \etab \detab + \etab+\etab = 2 \etab
\end{eqnarray*}
but
$$
\eta^2 \deta = \eta (\deta \eta +1) = ... = 2 \eta
$$
Thus e.g.:
\begin{eqnarray*}
\sp{\etab^2}{2+3\etab^2} &=& \eta^2\mbo
e^{\partial_{\eta} \partial_{\etab}}\mbox{ }
(2+3\etab^2) \mbo \vert_{\eta=0=\etab}\\
 &=& \eta^2 \sum_{r=0}^{\infty}
\frac{\partial_{\eta} \partial_{\etab}}{r!}
\mbo (2+3\etab^2)
\mbo \vert_{\eta=0=\etab}\\
 &=& 3 \eta^2 \frac{\deta^2\detab^2}{2}
\etab^2\mbo \vert_{\eta=0=\etab} = 6
\end{eqnarray*}
Since the scalar product formula Eq.\ref{spalt} relies on
ordinary commutative integration over the complex plane, it cannot
be used in the noncommutative case.
We will however be able to use a generalisation of Eq.\ref{spneu}
\footnote{It can be used even
in the fermionic case where one would usually apply
Berezin integration \cite{ak-lmp-bf}.}
on the noncommutative Bargmann Fock space \cite{ak-lmp-bf,ak-jmp-bf}.

We put the Bargmann Fock functions
of position eigenvectors Eq.\ref{bfdelta}
into Eq.\ref{spneu} and obtain now a new
formula for the transformation from Bargmann Fock to position space:
\be
\psi_x(x) = (2\pi L^2)^{-1/4} \left(
e^{-\frac{\eta^2}{2} + \eta\xl
-\frac{1}{4}(\xl)^2}\mbox{ }
e^{\partial_{\eta}\partial_{\etab}}\mbox{ }
\psi(\etab)\right)_{\eta=0=\etab}
\ee
and thus:
\be
\psi_x(x) = (2\pi L^2)^{-1/4} \left(
e^{-\frac{1}{2}\partial_{\etab}^2 + \xl \partial_{\etab}
-\frac{1}{4}(\xl)^2}\mbox{ }
\psi(\etab)\right)_{\etab=0}
\label{algebraischeruecktrafo}
\ee
The new transformation formula has the advantage that it can
be evaluated algebraically, i.e. by using the Leibniz rule to
commute the differentiation symbols to the right and
setting $\etab=0$.
Such techniques are
generalisable to the case of noncommutative $\etab$'s.
A similar calculation is possible for the transformation from
Bargmann Fock to momentum space.

\subsubsection{Pointwise multiplication}
When we will later abandon the position and
momentum spaces, which were classical manifolds,
we will still be able to obtain
all informations about
positions or momenta out of the matrix elements of the
position and momentum operators. However, the
operators $x$ and $p$ will not have eigenvectors
so that there will be nothing that could be called points in position
or momentum space. However, we will work with quantum field
theories in the path integral framework and
fields have to be multiplied pointwise in
position space in order to describe local interactions.
We thus have to translate the notion of pointwise
multiplication in position space, into Bargmann Fock space, before
we can generalise it.
To this end we use the formulas of the preceding subsection, which
connect the position space with the Bargmann Fock space:

The pointwise multiplication $*$ of functions on position space
translates into momentum space as the well known
convolution product:
\begin{eqnarray}
(\psi * \phi)(x) & =& \psi_x(x)\phi_x(x)\\
(\psi * \phi)(p) & =& (2\pi\hbar)^{-1/2} \int_{-\infty}^{+\infty}
dk\mbo \psi_p(k) \phi_p(p-k)
\label{conv}
\end{eqnarray}
To check this one takes two arbitrary functions on momentum
space, transforms them into position space, multiplies them pointwise
and transforms the resulting function back into momentum space. The
result is the convolution product formula Eq.\ref{conv}.

In order to derive the formula for pointwise multiplication
in Bargmann Fock space, we proceed the same way:
We transform two arbitrary Bargmann Fock functions into position space
using our new formula Eq.\ref{algebraischeruecktrafo},
multiply them pointwise and transform the result back into
Bargmann Fock space, using Eq.\ref{g1}:
\begin{eqnarray}
(\psi * \phi)(\etab) & = &
(2\pi L^2)^{-\frac{3}{4}} \int_{-\infty}^{+\infty} dx\mbo
e^{-\frac{1}{2} (\etab^2 + \partial^2_{\etab^{\prime}} +
\partial^2_{\etab^{\prime \prime}}) +
\frac{x}{L} (\etab + \partial_{\etab^{\prime}} +
\partial_{\etab^{\prime \prime}})
- \frac{3}{4} (x/L)^2 }
\psi(\etab^{\prime}) \phi(\etab^{\prime \prime}) \vert_0
\nonumber \\
 &   & \nonumber  \\
 & = & \left(\frac{2}{3\pi L^2}\right)^{\frac{1}{4}}
e^{\frac{1}{3}((\etab + \partial_{\etab^{\prime}} +
\partial_{\etab^{\prime \prime}})^2 -
\frac{1}{2} (\etab^2 + \partial^2_{\etab^{\prime}} +
\partial^2_{\etab^{\prime \prime}})}
\mbo \psi(\etab^{\prime}) \phi(\etab^{\prime \prime}) \vert_0
\label{star1dim}
\end{eqnarray}
Here and in the following $\vert_0$ will always stand for 'all
differentiations evaluated at zero'.
\sn
Let us also mention a 'weighted' pointwise multiplication that
we may call $*^{\prime}$:
\be
(\psi *^{\prime} \phi)(x) := \psi_x(x) \phi_x(x)\mbo
e^{\frac{1}{4} \left(\frac{x}{L}\right)^2}
\ee
In Bargmann Fock space this now takes a particularly
simple form:
\begin{eqnarray}
(\psi *^{\prime} \phi)(\etab) &=&
(2\pi L^2)^{-\frac{3}{4}} \int_{-\infty}^{+\infty} dx\mbo
e^{-\frac{1}{2} (\etab^2 + \partial^2_{\etab^{\prime}} +
\partial^2_{\etab^{\prime \prime}}) +
\frac{x}{L} (\etab + \partial_{\etab^{\prime}} +
\partial_{\etab^{\prime \prime}})
- \frac{1}{2} (x/L)^2 }
\psi(\etab^{\prime}) \phi(\etab^{\prime \prime}) \vert_0
\nonumber \\
 &=& (2\pi L^2)^{-\frac{1}{4}}
e^{\etab \partial_{\etab^{\prime \prime}} + (\etab +
\partial_{\etab^{\prime \prime}}) \partial_{\etab^{\prime}}}
\mbo \psi(\etab^{\prime}) \phi(\etab^{\prime \prime}) \vert_0
\nonumber \\
&=& (2\pi L^2)^{-\frac{1}{4}}
\psi(\etab + \partial_{\etab^{\prime}})\phi(\etab + \etab^{\prime})
\vert_0
\label{starp1dim}
\end{eqnarray}
Recall that the Bargmann Fock polynomials $\etab^m$
read in the position eigenbasis (Eq.\ref{unha1}):
\be
\sp{x}{\etab^m} =
\sqrt{m!}\mbo H_m (x)\mbo  e^{-\frac{1}{4}\left(\frac{x}{L}\right)^2}
\ee
where $H_m$ is the $m$'th Hermite polynomial.
The pointwise multiplication $*$ of these basis functions thus yields
a function of even more rapid decay at infinity. Their product
is decreasing like $ ~ e^{-\frac{1}{2}\left(\frac{x}{L}\right)^2}$.
On the other hand we see that the weighted multiplication
$*^{\prime}$ would cancel one of the Gau{\ss}ian factors and thus keep
the asymptotic behaviour unchanged under the multiplication.

For field theory we will later discuss the possibility to use, as a
technical trick, the simpler to handle $*^{\prime}$ instead of $*$.
Choosing $L$ large (recall that the choice of
$L$ is arbitrary and does not affect the theory),
we approximate $*$, except only for its infrared behaviour.

\subsection{Generalised quantum mechanics in the Barg\-mann Fock basis}
In the following four subsections we deal with generalisations
of quantum mechanics, which have been developed in
\cite{ak-lmp-bf,ak-jmp-bf,ak-jmp-ucr}. The Bargmann Fock
functions now become noncommutative. However they still span a
Hilbert space of states and still
operators can be represented as integral kernels.
On each such Hilbert space we will
represent a class of generalised Heisenberg
algebras of positions and momenta. We focus on two typical examples
and demonstrate how nonzero minimal uncertainties
 in positions and momenta arise
from the generalised commutation relations. The study of possible
generalisations of the notion of pointwise multiplication, which will
later be needed for the description of local interactions, is new.

\subsubsection{Noncommutative Bargmann Fock spaces}
In section 2.1 we recalled that the Heisenberg algebra of the $\x$
and $\p$ can be represented on functions on position space
$\sp{x}{\psi}$, on functions on momentum space $\sp{p}{\psi}$,
or e.g. on functions $\psi(\etab)$ on Bargmann Fock space.
The corresponding changes of the bases, i.e. the
unitary transformations that map functions on
Bargmann Fock space to functions on position or momentum
space, and vice versa, were given.

Obviously, a generalised Heisenberg algebra that implies
nonzero minimal uncertainties in positions and momenta,
cannot act on functions in a position or momentum eigenbasis,
 like $\sp{x}{\psi}$ or $\sp{p}{\psi}$. This is simply because
there will be no position or momentum eigenstates $\v{x},\v{p}$
in the
representations of the Heisenberg algebra. However, we will see
that such Heisenberg algebras can still be represented on
Hilbert spaces of now noncommutative Bargmann Fock functions
$\psi(\etab)$.

Our proceeding is as follows: We start by giving two examples
of noncommutative Bargmann Fock spaces. We discuss their properties
and show e.g. that operators on these
 spaces can still be written also as
integral kernels. Secondly then we find two
generalised Heisenberg algebras that can be represented on these
spaces. It is shown that these two Heisenberg algebras indeed
imply the existence of nonzero minimal
uncertainties in positions and momenta.
Thus, as we said,
there is then no position or momentum eigenbasis to which
 we could transform. We will discuss how all information
on positions and momenta can still be obtained, namely
from the matrix elements of the $\x_i$ and $\p_j$.
\mn
Explicitly, we consider e.g. the following
 noncommutative generalisation of
$n$- dimensional Bargmann Fock space:
The states $\v{\psi}$ are given as polynomials or power series
$\psi(\etab)$ where the coordinate functions $\etab_i$ and
$\eta_i$ are
now noncommutative and defined to obey $(q \ge 1)$, see \cite{ak-lmp-bf}:
\be
\label{1bfi}
\etab_a \etab_b - q \etab_b \etab_a = 0 \mbox{ \qquad for \qquad }a>b
\ee
\be
\eta_a \eta_b - q \eta_b \eta_a = 0 \mbox{ \qquad for \qquad }a<b
\ee
\be
\eta_a \etab_b - q \etab_b \eta_a = 0 \mbox{ \qquad for \qquad }a \ne b
\ee
\be
\eta_a \etab_a -  \etab_a \eta_a =
(q^2 - 1) \sum\limits_{i\le a} \etab_i \eta_i
\ee
The scalar product of two Bargmann Fock polynomials
or power series $\psi(\etab)$ and $\phi(\etab)$,
later to be interpreted as states $\v{\psi}, \v{\phi}$, is unique.
It can be calculated by a generalised formula Eq.\ref{spneu}:
\be
\sp{\psi}{\phi} = \overline{\psi(\etab)}\mbo
e_{1/q}^{\partial_{\eta_i}\partial_{\etab_i}}
\mbo \phi(\etab) \mbo \vert_{\eta=0=\etab}
\label{spgen}
\ee
Here, the $q$- exponential is defined as (each $i$ summed over)
\be
e_{1/q}^{\partial_{\eta_i}\partial_{\etab_i}}
= \sum_{r=0}^{\infty}\frac{(
\partial_{\eta_i}\partial_{\etab_i})^r}{[r]_{1/q}!}
\ee
where
$$
[r]_{1/q} = 1 + q^{-2} + q^{-4} + ... + q^{-2(r-1)} =
\frac{q^{-2r}-1}{q^{-2}-1}
$$
and
$$
[r]_{1/q}! = 1\cdot [2]_{1/q}\cdot [3]_{1/q}\cdot ... \cdot [r]_{1/q}
$$
The evaluation of the scalar product formula Eq.\ref{spgen}
is algebraic: The Leibniz rules, or
commutation relations, are used to
commute the $\detab$'s to the right and the $\deta$'s to the left.
One evaluates at $\etab = 0 = \eta$, i.e. all terms except the
constants are set zero. The remaining number is the value of the scalar
product. The Leibniz rules and other relations
can be written in a compact form:
(the $i,j$ summed over)
\be
\detapb{a} \etab_b - q R_{ib}^{aj} \etab_j \detapb{i} = \delta_{ab}
\qqq
\detapb{a}\detapb{b} - \frac{1}{q} R_{ab}^{ij} \detapb{j}\detapb{i} =0
\label{twoa}
\ee
\be
\detapb{a}\detapp{b} - \frac{1}{q} (R^{-1})_{bi}^{ja}
\detapp{j}\detab{i}=0
\qqq
\detapp{a}\detapp{b} - \frac{1}{q} R_{ba}^{ji} \detapp{j}\detapp{i} =0
\ee
\be
\eta_a \detapp{b} - q R_{ja}^{bi} \detapp{i} \eta_j = \delta_{ab}
\qqq
\detapb{a}\eta_b - q R_{ab}^{ij} \eta_j \detapb{i} = 0
\ee
\be
\etab_a \detapp{b} - q R_{ba}^{ji} \detapp{j}  \etab_i =0
\label{twog}
\ee
where: (the $e_i^j$ are matrix units)
\begin{equation}
R = q \sum_i e^i_i \otimes e^i_i + \sum_{i\ne j} e^i_i \otimes e^j_j +
(q-1/q) \sum_{i>j} e^i_j \otimes e^j_i
\label{1bff}
\label{surm}
\ee
This generalised Bargmann Fock formalism introduces only one
parameter $q$. The formalism was
derived in \cite{ak-lmp-bf} where
it was shown that it is a minimal generalisation
under certain consistency conditions like the invariance of $*$- structure,
Poincar{\'e} series and positivity of the norm,
see also \cite{ak-proc-berlin}. The algebra of the Bargmann Fock generators
is conserved under the action of the quantum group $SU_q(n)$ and invariant
under the conjugation operation\footnote{Note that $\bar{ }$ is an
anti algebra morphism, so that e.g.
$\overline{\detapb{i}\etab_j} = \eta_j \detapp{i}$ (we defined
the $\deta$'s as right derivatives)}.
Technically the $\etab,\eta,\detab,\etab$ generate a
Fun$SU_q(n)$-comodule algebra, which means that the linear
action of the $SU_q(n)$ respects the commutation
relations as well as the $*$- structure. Thus the quantum group consists
of linear quantum canonical transformations of the algebra.
\mn
Let us consider the above commutation relations
of the Bargmann Fock generators in the one dimensional case, where
they reduce to:
\be
\detab \etab - q^2 \etab \detab = 1 \qquad \qquad
\eta \deta - q^2 \deta \eta = 1
\ee
\be
\etab \deta - q^2 \deta \etab = 0 \qquad \qquad
\detab \eta - q^2 \eta \detab = 0
\ee
\be
\eta \etab -q^2 \etab \eta = 0
\qquad \qquad \deta \detab -q^2 \detab \deta = 0
\ee
Using $n$ commuting copies of it we arrive at a second $n$- dimensional
generalisation of the Bargmann Fock construction (all $q_i \ge 1$):
\be
\label{2bfi}
\detab_i \etab_i - q_i^2 \etab_i \detab_i = 1 \qquad \qquad
\eta_i \deta_i - q_i^2 \deta_i \eta_i = 1
\ee
\be
\etab_i \deta_i - q_i^2 \deta_i \etab_i = 0 \qquad \qquad
\detab_i \eta_i - q^2 \eta_i \detab_i = 0
\ee
\be
\eta_i \etab_i -q_i^2 \etab_i \eta_i =0
\qquad \qquad \deta_i \detab_i -q_i^2 \detab_i \deta_i =0
\label{2bff}
\ee
where Bargmann Fock generators of different indices commute. It
provides a simple example of a Bargmann Fock construction where
there is no quantum group symmetry of the commutation relations.
This generalised Bargmann Fock space is
parametrised by $n$ parameters $q_i$.

Our two examples are certainly not the most general possible
consistent generalisations of the construction of Bargmann Fock
Hilbert spaces. Let us however postpone the study of a more general case.
We will see that
the two Bargmann Fock spaces given above can already serve
as representation spaces for interesting generalised Heisenberg algebras.

In both cases the Hilbert space is spanned, as usual, by the
ordered polynomials
$\etab_1^{r_1} \etab_2^{r_2}\cdot ... \cdot \etab_n^{r_n}$ which
are found to be still orthogonal and do now have the norm:
\be
\sp{\etab_1^{r_1} \etab_2^{r_2}\cdot ... \cdot \etab_n^{r_n}}{
\etab_1^{r_1} \etab_2^{r_2}\cdot ... \cdot \etab_n^{r_n}}
= \prod_{i=1}^{n} [r_i]_q!
\ee
in the first case and for the second kind of generalised
Bargmann Fock space it is:
\be
\sp{\etab_1^{r_1} \etab_2^{r_2}\cdot ... \cdot \etab_n^{r_n}}{
\etab_1^{r_1} \etab_2^{r_2}\cdot ... \cdot \etab_n^{r_n}}
= \prod_{i=1}^{n} [r_i]_{q_i}!
\ee
One also proves that the Bargmann Fock operators $\detapb{i}$ and
$\etab_j$ that act on the functions $\psi(\etab)$ as
differentiation and multiplication\footnote{We hope it is
not too confusing if we denote the 'multiplication by $\etab_j$' operator
with the same symbol as the 'coordinate function' $\etab_j$. They do
of course obey the same commutation relations.} operators, are
still adjoint to each other in respect to the scalar product, i.e.:
\be
\sp{\etab_i \psi(\etab)}{\phi(\etab)} =
\sp{\psi(\etab)}{\detapb{i}\phi(\etab)}
\mbox{ \quad and \quad }
\sp{\detapb{i}\psi(\etab)}{\phi(\etab)} =
\sp{\psi(\etab)}{\etab_i\phi(\etab)}
\ee
This insures that position and momentum operators,
defined e.g. (like in ordinary quantum mechanics)
through linear combinations
$ x_j \propto (\etab_j + \detapb{j})$ and
$p_j \propto i(\etab_j - \detapb{j})$ are still symmetric operators
on our space of polynomials in $\etab$'s.

Before we study the position and momentum
operators in more detail, let us mention how, just like in
ordinary quantum mechanics, operators can be represented as
integral kernels.
\subsubsection{Integral kernels}
It is still
possible to represent operators $P$ on Bargmann Fock functions,
expressed in terms of
multiplication and differentiation operators
$\bar{\eta}_j$ and ${\partial}_{\bar{\eta}_j}$,
as integral kernels.
Once the operator $P(\etab,\detab)$ is normal ordered,
there is a simple rule how to get its integral kernel $G_P$,
which is a function of
${\bar{\eta}}^{\prime}$ and $\eta$. Integrating
any Bargmann Fock function
${\psi}(\bar{\eta})$ over $G_P({\bar{\eta}}^{\prime},{\eta})$
leads then to a
function of ${\bar{\eta}}^{\prime}$, which is
$P{\psi}({\bar{\eta}}^{\prime})$:
\begin{equation}
\int d{\bar{\eta}} d{\eta}\mbo
 G_P({{\bar{\eta}}}^{\prime},{\eta})\mbo
e_{1/q}^{{\partial}_{{\eta}_i}{\partial}_{\bar{\eta}_i}}
\mbo {\psi}({\bar{\eta}})
= P{\psi}({\bar{\eta}}^{\prime})
\end{equation}
Here, the integration is meant to be our algebraic scalar product, which
expresses the integration in terms of derivatives, i.e.
we defined for notational convenience:
\be
\int d{\bar{\eta}} d{\eta}\mbo  \overline{\psi(\etab)}\mbo
 e_{1/q}^{{\partial}_{{\eta}_i}{\partial}_{\bar{\eta}_i}}\mbo
{\phi}({\bar{\eta}})
 := \overline{\psi(\etab)}
\mbo e_{1/q}^{{\partial}_{{\eta}_i}{\partial}_{\bar{\eta}_i}}
\mbo \phi(\etab)\mbo \vert_{\eta=0=\etab}
\ee
For this to work,
the appropriate commutation relations between two copies (e.g. primed and
unprimed) of the function
space had to be calculated, see \cite{ak-jmp-bf}.
\sn
If we define e.g. the position operator as
$x_i := L_i (\etab_i + \detapb{i})$, it has
the integral kernel
\be
G_x({{\bar{\eta}}^{\prime}},\eta ) =
L\mbo (\etab_i^{\prime}\mbo
e_{1/q}^{{{\bar{\eta}}^{\prime}}_j{\eta}_j} +
\mbo e_{1/q}^{{{\bar{\eta}}^{\prime}}_j{\eta}_j}\mbo \eta_i )
\ee
and another example is the $n$- dimensional
isotropic harmonic oscillator
$H := \omega \etab_i \detapb{i}$ (summed over $i$).
Since $H$ is hermitean, the time evolution operator $U = e^{-i(t_f-t_i)H}$
is unitary. The eigenvalues of $H$ are:
\be
H\mbo \v{\etab_1^{r_1} \etab_2^{r_2} \cdot ... \cdot \etab_n^{r_n}}
= \omega [r_1+r_2+...+r_n]_q\mbo
\v{\etab_1^{r_1} \etab_2^{r_2} \cdot ... \cdot \etab_n^{r_n}}
\ee
where $[r]_q$ is the '$q$- number', i.e. the partial
geometric series:
\be
[r]_q := 1 + q^2 + q^4
+ ... + q^{2(r-1)} = \frac{q^{2r}-1}{q^2-1}
\ee
The integral kernel of $U$, i.e. the Greens function is then
found to be \cite{ak-jmp-bf}:
\be
G_U =
\sum_{r=0}^{\infty}
\frac{ { ( {{\bar{\eta}}^{\prime}}_i{\eta}_i)}^r}{[r]_{1/q}!}
e^{-i\omega (t_f-t_i)[r]}
\label{easy gf}
\ee
These are the results for our first kind of generalised
Bargmann Fock construction, using the
relations Eqs.\ref{1bfi}-\ref{1bff}.
The corresponding formulas for our second example of a noncommutative
Bargmann Fock space (Eqs.\ref{2bfi}-\ref{2bff})
are easily derived from
the one-dimensional case above. An important difference is that the
eigenvalues of the operator $H := \omega \etab_i \detapb{i}$ then read:
\be
H\mbo \v{\etab_1^{r_1} \etab_2^{r_2} \cdot ... \cdot \etab_n^{r_n}}
= \omega ([r_1]_{q_1}+[r_2]_{q_2}+...+[r_n]_{q_n})\mbo
\v{\etab_1^{r_1} \etab_2^{r_2} \cdot ... \cdot \etab_n^{r_n}}
\ee
\subsubsection{Positions and momenta}
We now study the Heisenberg algebras of positions and momenta
that can be represented on our generalised Bargmann Fock spaces.
Let us start with the Heisenberg algebra that is generated by
operators $\x_r, \p_r,  (r=1,...,n)$ which are represented, just
like in usual quantum mechanics, as
\begin{equation}
\x_r = L_r (\etab_r + \detapb{r}) \mbox{ \qquad and \qquad }
\p_r = i K_r (\etab_r - \detapb{r})
\label{xpansatz}
\end{equation}
on our first Bargmann Fock space. Note that
since the $\detapb{i}$ and the
$\etab_j$ do not carry units, the newly
introduced constants $L_r$ and $K_r$ do.
\sn
The commutation relations of this Heisenberg
algebra then read as follows:
\begin{equation}
[\x_r,\p_r] = i \hbar + i \hbar (q^2 -1)
\sum_{s\le r}
\left( \frac{q^2+1}{2}\right)^{s-1}
\left(\frac{\x_s^2}{4 L_s^2}+\frac{\p_s^2}{4 K_s^2} \right)
\label{xpcr}
\label{1Hi}
\end{equation}
The mixed commutation relations read for $s>r$:
\begin{equation}
[\x_s,\p_r] =  -i \frac{K_r}{L_r} \frac{q-1}{q+1} \{ \x_s, \x_r\}
\qqq
[\x_s,\x_r] =  -i \frac{L_r}{K_r} \frac{q-1}{q+1} \{ \x_s, \p_r\}
\end{equation}
For $s<r$ one gets:
\begin{equation}
[\x_s,\p_r]  =  i \frac{L_s}{K_s} \frac{q-1}{q+1} \{ \p_s, \p_r\}
\qqq
[\p_s,\p_r]  =  -i \frac{K_s}{L_s} \frac{q-1}{q+1} \{ \x_s, \p_r\}
\label{1Hf}
\end{equation}
In ordinary quantum mechanics where $[\x,\p]=i\hbar$
the $L_i$ and $K_j$ must obey $L_rK_r = \hbar/2$,
in order to get exactly the term $i\hbar$ on the lhs.
In our generalised situation, in order to get in the commutation
relations Eq.\ref{1Hi} also exactly the term $i\hbar$, the $L_r$
and $K_r$ must obey
\begin{equation}
L_r K_r := \frac{\hbar}{2} \left(\frac{q^2 +1}{2}\right)^r
\label{klpbez}
\end{equation}
That the constant term on the rhs is exactly $i\hbar$ without any
$q$-factors, insures that
the generalised framework describes
ordinary quantum mechanical behaviour in medium scales,
see \cite{ak-jmp-ucr}. Recall that we want to develop generalisations
that deviate from ordinary quantum theory only in the very small and large
scales.

Let us also consider our second Bargmann Fock space. Using again
the usual
definitions Eq.\ref{xpansatz} we can represent on it the
 Heisenberg algebra with the commutation relations:
\begin{equation}
[\x_i,\p_j] = i \hbar \delta_{ij} + i \hbar \delta_{ij} (q^2-1)
 \left(\frac{1}{4L_i^2} \x_i^2 + \frac{1}{4 K_i^2} \p_i^2
\right)
\label{2Hi}
\end{equation}
\be
[\x_i,\x_j] = 0
\ee
\be
[\p_i,\p_j] = 0
\ee
where
\be
L_i K_i = \hbar (q^2 +1)/4
\label{2Hf}
\ee
Note that if $q=1$, or for the second
Heisenberg algebra if all $q_i=1$,
the constants $K_i$ and $L_i$ drop out of
the commutation relations. This reflects
that in ordinary quantum mechanics a length or
a momentum scale can only be set
by the Hamiltonian i.e. by choosing a particular system.
Here, for $q > 1$, or some $q_i>1$, the
$K_r$ and $L_s$ do appear in the commutation relations.
These scales now become a property
of the quantum mechanical formalism itself. Thus, although the
generalisation parameters $q_i$ of
the Bargmann Fock spaces are unitless, there do now appear length and
momentum scales in the framework.
\mn
As we said, one can show that our generalised Heisenberg algebras
describe ordinary quantum
mechanical behaviour in the physical region of not too large
and not too small positions and momenta, explicitly where:
\begin{equation}
 (\Delta x_{0})^2 \ll \langle \x^2\rangle \ll
 \frac{{\hbar}^2}{4(\Delta p_{0})^2}
\mbox{ \qquad and \qquad }
(\Delta p_{0})^2 \ll \langle \p^2\rangle \ll
 \frac{{\hbar}^2}{4(\Delta x_{0})^2}
\end{equation}
This is nicely consistent with the
wave-particle dualism, because it means that the critical upper
length scale, where the formalism begins to deviate significantly,
is just the length needed to measure momentum
to its minimal uncertainty. Similarly,
the critical upper momentum is just the
momentum needed to measure the position to its minimal uncertainty.
Note that the existence of this region of effectively ordinary
quantum theoretical behaviour relies on the fact that in our generalised
Heisenberg algebras the constant term
on the rhs is exactly $i\hbar \delta_{ij}$
without any $q$- factors \cite{ak-jmp-ucr}.
\mn
The functional analysis of our position and momentum operators is
as follows \cite{ak-jmp-bf}:
We start by defining their domain $D$ to be the set of
only those Bargmann Fock functions that are polynomials in $\etab$'s.
This is a dense set in the Hilbert space and on it the $\x_i$ and
$\p_i$ are symmetric with their ranges in the domain.
The $\x_i$ and $\p_i$ are no longer essentially self-adjoint.
Their adjoints $\x_i^*$ and $\p_i^*$ are closed but nonsymmetric.
The $\x_i^{**}$ and $\p_i^{**}$ are closed and symmetric, but
their deficiency indices
no longer vanish. The deficiency subspaces are still of the
same size so that there are continuous
families of self adjoint extensions.

One might be tempted to try to fix the choice of the self
adjoint extensions by the requirement that the domains coincide. A
diagonalisation of the $\x_i$ or the $\p_i$ would then lead as usual
to momentum space or position space representations.
However, one can show that there are no
more eigenstates in the
representations of the Heisenberg algebras, and there is thus no
position or momentum eigenbasis.

The mechanism is the same for both of our examples of
generalised Heisenberg algebras. It can already be seen in the
one-dimensional case that non of the observables $\x$ or $\p$ is
diagonalisable on a representation of the
 Heisenberg algebra, nor does it
have any eigenvectors: \sn
The commutation relation
\begin{equation}
[\x,\p] = i \hbar + i \hbar (q^2-1)
\left(\frac{1}{4L^2} \x^2 + \frac{1}{4 K^2} \p^2
\right)
\label{1dimcr}
\end{equation}
leads to the uncertainty relation:
\begin{equation}
\Delta x \Delta p \ge \frac{\hbar}{2} \left( 1 + (q^2-1)\left(
\frac{(\Delta x)^2 + \langle \x\rangle ^2}{4L^2}
+ \frac{(\Delta p)^2 + \langle \p\rangle ^2}{4K^2}\right)
\right)
\label{ucr1}
\end{equation}
It implies nonzero minimal uncertainties in $\x$ as well as $\p$
measurements. This can be seen as follows: As e.g.
$\Delta x$ gets smaller, $\Delta p$ must increase so that
the product $\Delta x \Delta p$ of the lhs remains larger than the
rhs. In usual quantum mechanics this is always possible, i.e.
$\Delta x$ can be made arbitrarily small. However, in the
generalised case, where $q>1$, there is a $(\Delta p)^2$ term on the
rhs which eventually grows faster with $\Delta p$ than the lhs.
Thus $\Delta x$ can no longer become arbitrarily small. The
minimal uncertainty in position measurements comes out as:
\begin{equation}
\Delta x_{0} = L \sqrt{1-q^{-2}}
\ee
Analogously one obtains the smallest uncertainty in the
momentum:
\be
\Delta p_{0} = K \sqrt{1-q^{-2}}
\end{equation}
where we have from Eq.\ref{klpbez} that $K=(q^2+1)\hbar /4L$.
For the details on the functional analysis and the uncertainty
relation see \cite{ak-jmp-ucr}, also for the $n$ dimensional case which
is analogous.
\sn
The essential point is, that due to the nonvanishing minimal
uncertainties in positions and momenta, there can no longer be
$\x$ or $\p$ eigenstates in the representations of the Heisenberg algebra.
To see this, recall that if we had
an $\x$ eigenstate $\v{x_0}$ with $\x.\v{x_0}=x_0\v{x_0}$ it would
have no uncertainty in the position measurement
\be
(\Delta x)^2_{\v{x_0}} =
\langle x_0\vert (\x - \langle x_0\vert
\x \vert x_0 \rangle)^2 \v{x_0} = 0
\ee
which would be a contradiction. There are thus no physical states
which are eigenstates of $\x$ or $\p$. We should clarify how we
identify the space of physical states, which is actually
smaller than e.g. our Hilbert space of
(weighted) square integrable Bargmann Fock functions: The space of
physical states has of course to be a representation space
on which the expectation values of positions and momenta are well defined.
Mathematically this means that it has to be a
representation of the Heisenberg algebra (i.e. not
of $\x$ or of $\p$ alone!)
on which both, $\x$ and $\p$ are symmetric and have their ranges in
the domain. The space $D$ of Bargmann Fock polynomials
is such a physical domain.
However, its completion with respect to the ordinary norm
is too large to be a representation of the Heisenberg algebra.
It should be
interesting to find a suitable norm with respect to which
our physical domain $D$ could be completed to a domain $\bar{D}$,
which would be a representation of the Heisenberg algebra, i.e.
on which $\x$ and $\p$ would remain symmetric while also
having their range
in the domain.
\sn
We will see how the nontrivial functional analytic
 behaviour of $\x$ and $\p$,
i.e. their non essentially-self-adjointness manifests
 itself also in the
physical description of positions and momenta.
\mn
Recall that all information on positions and momenta is
 contained in the
matrix elements of the position and momentum operators, and
matrix elements can of course be calculated in any basis.
In our Bargmann Fock basis matrix, elements of e.g. the
position
operators are calculated as
\be
\me{\psi}{\x}{\phi} = \overline{\psi(\etab)}\mbo
e_{1/q}^{\deta\detab}\mbo L(\etab + \detab)\mbo \phi(\etab)
\mbo \vert_{0}
\ee
In particular, we can calculate Bargmann Fock functions
 that denote states
$\psi_{(x^{\prime})}(\etab)$,
which approximate position eigenstates to
the eigenvalue $x^{\prime}$. To this end we start with the
eigenvalue equation:
\be
\x.\v{\psi_{(x^{\prime})}
} = x^{\prime} \v{\psi_{(x^{\prime})} }
\mbox{ \qquad i.e. \qquad } L(\etab + \detab)
\psi_{(x^{\prime})}(\etab)
= x^{\prime} \psi_{(x^{\prime})}(\etab)
\ee
Expanding $\psi_{(x^{\prime})}(\etab)$
 in the $\etab$'s yields a recursion
formula for the coefficients of the expansion.
 In ordinary quantum mechanics
the solution is the Dirac $\delta$ 'function', transformed into
Bargmann Fock space, i.e. Eq.\ref{bfdelta}
(with $x^{\prime}$ instead of
$x_0$).
In the generalised setting the (infinite)
recursion can also be solved, but now the power series that solves it
is no longer contained in the representation of the Heisenberg
algebra, since this would of course contradict the uncertainty relation.
However, every polynomial approximation of the power series is contained
in the above defined domain $D$. It is thus contained in
the representation of the Heisenberg algebra, and is therefore a
physical state. Using such states
we can still describe very localised particles.
We can even use approximating polynomials of
arbitrarily high degree, and thus come 'nearer and nearer' to
the above described power series that would be a position eigenvector.
However, and this is now an important effect of the non essential
self-adjointness of our $\x$ and $\p$- operators, no
polynomial describes a localisation that is more precise than
$\Delta x_0$. Similarly, in this way we can also calculate states
that describe particles of relatively
precise momentum, however of course
not more precise than $\Delta p_0$.

In contrast to the $\delta$- 'functions' of ordinary
quantum theory, these approximating states are
no longer orthogonal to each other (not even the unphysical power series
that would solve the eigenvalue equation are orthogonal).
For more details and a graph of their scalar
product see \cite{ak-jmp-ucr}. It should also be interesting
to study whether a generalised fourier transformation can be
found that would allow to easily transform information on positions
into information on momenta. In this context, compare also with the
generalised quantum mechanics (with discrete $x$ and $p$ spectra)
developed in \cite{toy1,toy2}, where techniques developed in
\cite{kw} allow such generalised fourier transformations.
\mn
To summarise, we arrived at the following picture:
\sn
While in classical mechanics the states can
have exact positions and momenta, in quantum mechanics there is the
uncertainty relation that does not allow $\x$ and $\p$ to have common
eigenvectors. Nevertheless
$\x$ and $\p$ separately do have 'eigenvectors',
though non-normalisable ones. The
spectrum is continuous, namely the whole configuration or momentum
space.

Our 'noncommutative
geometry'- or quantum group generalisations of the Heisenberg algebra
have further consequences for the observables $\x$ and $\p$:
It is not only that the $\x$ and $\p$
have no common eigenstates.
The uncertainty relation now implies that they even have no
eigenvectors at all in the representation
of the Heisenberg algebra.
This means of course the non-existence of absolute precision in
position or momentum measurements. Instead there are
nonzero minimal uncertainties in these measurements.
\bn
Let us discuss the level of generalisation that we
 have achieved so far.
\sn
The two generalised Heisenberg algebras
(Eqs.\ref{1Hi}-\ref{1Hf} and Eqs.\ref{2Hi}-\ref{2Hf}) that we
have studied are special examples of Heisenberg
algebras with commutation relations of the
form (summed over repeated indices):
\be
[\x_r,\p_s]  =  i\hbar \delta_{rs} + i\hbar A_{rs,tu} \{ \x_t,\x_u\}
+ i\hbar B_{rs,tu} \{\p_t,\p_t\}
\label{genHi}
\ee
\be
[ \x_r ,\x_s]  =  i C_{rs,tu} \{ \x_t, \p_u \}
\ee
\be
 [\p_r,\p_s]  =  i D_{rs,tu} \{ \x_t, \p_u \}
\label{genHf}
\ee
where the $A,B,C,D$ are real matrices, i.e. all
$A_{rs,tu}\in \bf R \rm$ etc.\newline
We now know that e.g. if the diagonals of the matrices
$A$ and $B$ are positive,
which is e.g. the case in our examples, this leads
 to the appearance
of nonzero minimal uncertainties in momenta and
in positions respectively.
\sn
We saw in particular how such a matrices $B$ can
introduce a term proportional to $(\Delta p)^2$
into the uncertainty relations. As we will
be discussing later in this paper, exactly this type
of correction to the uncertainty relation has actually already been
suggested to appear at small scales like the Planck scale.
\sn
On the other hand we saw how the introduction of a matrix $A$
leads to the appearance of
nonzero minimal uncertainties in momentum. As we said, we expect that
ordinary large scale gravity, i.e. curvature, causes such
uncertainties in momentum. Therefore it has to be shown that the momentum
operators, defined as the generators of translations, obey commutation
relations like those obtained from the introduction of a matrix $A$, when
working on a space which is flat only up to first order corrections.
A forthcoming paper on this subject is in progress.
\mn
In the present paper we are however
merely interested in quantum theory alone and its generalisability.
We were naturally led to consider Heisenberg algebras of the form
of Eqs.\ref{genHi}-\ref{genHf}. We dealt with
two examples of Heisenberg algebras
of this general form which we represented on our
$q$ - Bargmann Fock spaces. These Heisenberg algebras
have $n$ length scales $L_i$ and $1$ dimensionless parameter $q$ or
$n$ dimensionless $q_i$'s respectively.
However, let us note that more Heisenberg algebras of the
form \ref{genHi}-\ref{genHf} can be represented on our $q$- Bargmann
Fock spaces. This is because,
whenever we have a representation of
a Heisenberg algebra specified by matrices $A,B,C,D$, we can on the
same space also represent an entire class of Heisenberg algebras.
This class consists of the Heisenberg algebras specified by matrices
$A^{\prime},B^{\prime},C^{\prime},D^{\prime}$, where
\begin{eqnarray}
A^{\prime}_{ab,cd} & = & M^{-1}_{ai} M_{jb} M_{kc} M_{ld} A_{ij,kl}
\\
B^{\prime}_{ab,cd} & = & M^{-1}_{ai} M_{bj} M^{-1}_{ck}
 M^{-1}_{dl} B_{ij,kl}
\\
C^{\prime}_{ab,cd} & = & M^{-1}_{ai} M^{-1}_{bj} M_{kc}
 M^{-1}_{dl} C_{ij,kl}
\\
D^{\prime}_{ab,cd} & = & M_{ia} M_{jb} M^{-1}_{ck}
 M_{ld} D_{ij,kl}
\end{eqnarray}
for arbitrary $M\in GL(n,\bf R\rm)$. To see this,
 note that all Heisenberg
algebras within one class are isomorphic with the
isomorphism given on the
generators as
\be
\x_r \rightarrow \x_r^{\prime} = M^{-1}_{rs} \x_s \qquad
\quad \p_r \rightarrow \p_r^{\prime} = M_{sr} \p_s
\ee
Nevertheless, since the actual commutation relations
between the position
and momentum operators are not invariant, these Heisenberg algebras all
describe different physical behaviour. Note in particular that the
functional analysis of the $\x_i$ and $\p_j$ changes drastically when
there are negative
 diagonal elements of the matrices $A$ and $B$, (see
\cite{ak-jmp-ucr}). This case deserves further study.

To summarise, we can now represent two classes of generalised
Heisenberg algebras on $q$ Bargmann Fock Hilbert spaces. They will
serve us as examples for our studies of the influence of nonzero minimal
$\x$ and $\p$- uncertainties on the infrared and ultraviolet
behaviour in quantum field theories.
We postpone the explicit construction of the Hilbert space
representations of generalised Heisenberg algebras that are characterised by
fully arbitrary real matrices $A,B,C,D$. The appearance of nonzero minimal
position and momentum uncertainties then follows from the
uncertainty relations, e.g. in the case of positive diagonals of the
matrices $A$ and $B$.

\subsubsection{Pointwise multiplication}
The notion of pointwise multiplication is used for the description
of local interactions in quantum field theory. In our
generalised framework there are no more position eigenstates, i.e.
there are no more 'points in position space'. One must now develop
a generalised notion of 'noncommutative locality'.
\mn
In ordinary quantum mechanics, states $\v{\psi}$ can be expanded in the
position eigenbasis $\{ \v{x} \}$ to obtain wave functions on
position space $\psi_x(x) = \sp{x}{\psi}$. Their pointwise multiplication
\be
(\psi * \phi)(x) := \psi_x(x) \phi_x(x)
\ee
is the convolution product in momentum space and it reads
in Bargmann Fock space for $n$ dimensions (see Eq.\ref{star1dim}):
\be
(\psi * \phi)(\etab)
= \left(\frac{2}{3\pi L^2}\right)^{\frac{n}{4}}
e^{\sum_{i=1}^{n} \frac{1}{3}((\etab_i + \partial_{\etab_i^{\prime}} +
\partial_{\etab_i^{\prime \prime}})^2 -
\frac{1}{2} (\etab_i^2 + \partial^2_{\etab_i^{\prime}} +
\partial^2_{\etab_i^{\prime \prime}})}
\mbo \psi(\etab^{\prime}) \phi(\etab^{\prime \prime}) \vert_0
\label{starndim}
\ee
The weighted pointwise multiplication
$(\psi *' \phi)(x) := \psi_x(x) \phi_x(x)
e^{\frac{x_ix_i}{4 L^2}}$
reads in $n$ dimensional Bargmann Fock space simply,
see Eq.\ref{starp1dim}:
\be
(\psi *^{\prime} \phi)(\etab) =
(2\pi L^2)^{-\frac{n}{4}}
\psi(\etab + \partial_{\etab^{\prime}})\phi(\etab + \etab^{\prime})
\vert_0
\label{starpndim}
\ee
These formulas are obtained by taking two functions on Bargmann
Fock space, transforming them into position space, multiplying them
pointwise and then transforming the result back into Bargmann Fock space.

An analogous proceeding is of course not possible in our generalised
quantum mechanics. The position operators $\x_i$ do no longer have
eigenvectors in the representation of the Heisenberg algebra. We abandoned
the position eigenbasis and thus also
functions on position space which could
be multiplied pointwise.
We succeeded in abandoning the notion
 of points, but what should now be the
generalised notion of 'noncommutative locality'? For the pointwise
multiplication one may be tempted to simply stick to Eq.\ref{starndim},
even for $q>1$. However, writing the exponential as a power series
there are terms like e.g. $\etab_i^2\etab_j^2$ or equally
$\etab_j^2\etab_i^2$. In the generalised framework
such terms are no longer
equal, due to the noncommutativity. There is thus no unique
generalisation of Eq.\ref{starndim}.

Clearly the noncommutative generalisation of the notion of pointwise
multiplication, i.e. of locality, has to be developed on physical
grounds. This is a very crucial point in our generalisation of
quantum theory and we will here not try to give a general answer.
Let us however note, that the weighted pointwise
 multiplication $*^{\prime}$
does appear to be straightforwardly generalisable since it does not
involve any functions other than the two Bargmann Fock functions
that are to be multiplied. In spite of its
modified infrared behaviour
we will later use $*^{\prime}$ as a working hypothesis.

\section{Quantum field theory with nonzero minimal uncertainties}
We will be working with the path integral approach to quantum
field theory. Our aim is to generalise its Heisenberg algebra
of 4- positions and 4- momenta.
In quantum field theory one would usually focus on the
Heisenberg algebra of the fields and their conjugate momentum fields.
Let us therefore clarify the r{\^o}le of
this Heisenberg algebra of the positions and the momenta,
which we are going to generalise.
\mn
In quantum field theory too there is
a Heisenberg algebra generated by the $\x_{\mu}$ and $\p_{\nu}$,
which, in the euclidean case, obey the commutation relations
$[\x_{\mu},\p_{\nu}] = i\hbar \delta_{\mu\nu}$.
This Heisenberg algebra is represented on the (Hilbert) space of
fields $\phi$. The quantum field theoretical
partition function $Z=\int D\phi\mbo e^{iS[\phi]/\hbar}$ is the sum over
all amplitudes $e^{iS[\phi]/\hbar}$ where $S$ is the action functional
and the e.g. bosonic fields $\phi$ run through the representation
space of the Heisenberg algebra.
\mn
A 'field on position space' is given by the
coordinates $\phi(x)$ of the abstract field $\phi$
in the position eigenbasis.
In this basis the scalar product of fields reads
$(\phi(x),\phi^{\prime}(x)) = \int d^4x
\mbo \phi(x) \phi^{\prime}(x)$. The scalar product
of momentum and position eigenvectors $\phi_{p}$ and $\phi_{x}$, is
$(\phi_p,\phi_x) = (2\pi\hbar)^{-2} e^{-ix.p/\hbar}$. Therefore the
unitary transformation
to the field's coordinates in the momentum eigenbasis reads:
$\phi(p) = \int d^4x\mbo (\phi_p,\phi_x) \phi(x)$.
\mn
Analogous to the situation in
quantum mechanics, the position and momentum operators are
essentially self-adjoint with (non-normalisable) 'eigenvectors', namely
plane waves or Dirac $\delta$- peaks.
These eigenvectors still\footnote{
This is analogous to the situation in quantum mechanics.
Positions and momenta do not become mere 'parameters'
in quantum field theory.} are the 'points' in position or in
momentum space, which means that e.g. the coordinates $p_{\nu}$
of a 'point' in momentum space are eigenvalues of the operators $\p_{\nu}$.
\mn
Choosing an
arbitrary positive length $L$ we define Bargmann Fock operators
$\etab_{\mu} := \frac{1}{2L} \x_{\mu}  -\frac{iL}{\hbar} \p_{\mu}$ and
$\detab_{\mu} := \frac{1}{2L} \x_{\mu}  +\frac{iL}{\hbar} \p_{\mu}$.
Fields can then be written as Bargmann Fock functions $\phi=\phi(\etab)$.
Recall that $\phi(\etab)$ denotes the
field obtained by applying
the function $\phi(\etab)$ of the Bargmann Fock operators $\etab$
on that field which is eigenvector to all $\detab_{\mu}$ with the
eigenvalue $0$ (a gau{\ss}ian), see end of Sec.\ref{cadi}.
\mn
Also, arbitrary fields $\phi$ can be expanded
in the countable eigenbasis of the operator
$\etab_{\mu}\detab_{\mu}$ to obtain the field's
coordinates $\phi_{r_1,r_2,r_3,r_4}$
in this Bargmann Fock basis.
Similarily, the operators on the fields can
be expanded in the Bargmann Fock basis. The scalar product
reads $(\phi,\phi^{\prime})=\sum_{r_1,r_2,r_3,r_4}^{\infty}
\phi_{r_1,r_2,r_3,r_4} \phi^{\prime}_{r_1,r_2,r_3,r_4}$ in this basis.
\mn
The path integral is usually written in the position
eigenbasis, which means to work with with the field's
coordinates $\phi(x)$ in the position eigenbasis. The functional $S$
is then also written in the position eigenbasis i.e. it is
expressed in terms of integrals and operators
on functions over position space. It is often
more convenient, e.g. for the derivation of Feynman rules,
to consider the path integral, i.e. the same abstract
fields $\phi$ and the same
abstract action functional $S$ in the momentum eigenbasis instead.
The action functional $S$ is then expressed in terms
of integrals and operators on functions over momentum space.
\mn
Similarly, and this will be our choice, we can consider the
path integral, i.e. the same
abstract fields $\phi$ and the same abstract action functional $S$
in the Bargmann Fock basis of the function space.
Before we further discuss quantum field theory in
the Bargmann Fock basis, let us come back to
the (of course basis independent)
meaning of the operators $\x_{\mu},\p_{\nu}$ and the fields $\phi$ that
they are acting on.
\mn
In relativistic quantum field theory,
the r{\^o}le of the 4-position and 4-momentum operators,
the fields and their propagators
is not as simple as in quantum mechanics.
The coordinates $\phi(x)$ of a field $\phi$
in the position eigenbasis are not the 'probability amplitude
for finding a particle at $x$', and Greens functions $G_2(x_2,x_1)$
do not simply propagate particles from $x_1$ to $x_2$:
It is not sufficient to know the amplitudes
on a given spacelike surface to be able to calculate the amplitude
at some point $P$ in the future, even when neglecting interactions.
Problems with Newton Wigner localisation-
 and propagation are related to that,
see e.g. \cite{rui}.
The reason for this is the unavoidable existence of antiparticles:
\mn
What has to be known in order
to calculate the field's amplitude at a point $P$ in space-time
is not only the amplitudes of positive
frequencies on an earlier spacelike surface but also the
amplitudes of negative frequencies on a later spacelike surface,
since they both propagate towards $P$, see e.g. \cite{feynman-qed}.
This is to be compared with the
relativistic propagation of an ordinary measurable field,
e.g. a classical electrical potential or a directly measurable
probability wave. One would then apply
retarded boundary conditions to the equation of motion.
This would ensure causal propagation of the wave (with the
propagator vanishing for spacelike distancies). The knowledge of the
field on a spacelike surface would be sufficient to calculate its
values at later times.
\mn
However, in quantum field theory, not the fields are measurable. It is the
particles, or more precisely, it is
the expectation values of their 4-positions and
4-momenta that are,
in principle\footnote{When speaking of 'in principle possible'
measurements we are here and in the following not
distinguishing virtual and real particles. Let us omit the
discussion that for actual measurements on real particles
one would consider sufficiently well separated wave packets
of in- or out-going particles, each on the mass shell, the
S-matrix limit etc.}, measurable.
Therefore the boundary conditions
for the propagators must be such that
the particles propagate causally, rather than the unobservable 'waves'.
This makes a difference because
while the positive energy components, i.e. the particles
are of course to be propagated forward in time, i.e. retarded,
 the negative
energy components, i.e.
the anti-particles components are to be propagated backwards in
time, i.e. with advanced boundary conditions.
\mn
As was pointed out by Feynman \cite{feynman in dirac lecture},
it is not
difficult to see that it is relativity which
does not allow a restriction to positive frequency components alone:
Let us transform the propagator of the positive frequencies only,
into the time coordinate. A theorem in fourier theory
then applies, which
tells that the fourier transform
of a function that is nonzero only on a half axis, cannot vanish on
any finite interval.
\mn
Therefore, the propagation of positive
frequencies cannot vanish for any finite time interval.
In particular, positive frequencies are also propagated
 between events,
say E1 and E2, that are spacelike separated. In a suitable chosen
Lorentz transformed frame the events E1 and E2 occur in the
reversed order. In this frame, negative frequencies
propagated backwards in time.
This explains why relativity requires antiparticles, and why
the antiparticle's properties are determined by the
properties of the
particles.
\rm
We saw on the other hand that
since not all frequencies can be propagated forward
in time, propagation
also in spacelike distances is unavoidable. E.g. for free massive
particles there is an exponentially decaying propagation
outside the lightcone. One might be surprised about this result
since from the operator approach to quantum field theory
 one knows that
spacelike separated field operators commute and also, naivly
interpreted, it seems to contradict causality.
\mn
Recall however that the fields that are propagated by the
quantum field theoretical propagators are not directly observable.
In particular, space-time and momentum-energy are not classical
parameter spaces. The information on space,
time, momentum and energy is
encoded in the formal expectation values of the corresponding
operators in the Heisenberg algebra. These
expectation values obey the
uncertainty relations that are implied by the
commutation relations of the
position and momentum operators $\x_{\mu}$ and $\p_{\nu}$.
\mn
In particular, there is a time-energy uncertainty relation.
More precisely,
the energy- and the time coordinates are related by a
 unitary mapping,
which is found to be a fourier transformation. Any
restriction, i.e. any
knowledge on the energy, automatically leads to consequences, or
'uncertainties' in $x_0$, i.e. in time. We saw an example
 of exactly
this above, namely
that propagation
restricted\footnote{More precisely, restricted
to energies larger or equal to the restmass of the particle}
to positive frequencies
implies nonzero propagation to all finite time intervals, and hence
also to spacelike distances. The reason why e.g. the propagator
for massive particles cannot vanish outside the lightcone (but has
an exponential tail instead) is therefore the following:
\mn
In order to decide whether particles propagate outside the
lightcone one measures not only their expectation
 values of position and time.
One also measures whether a particle has arrived at all, i.e. one
measures also whether positive energy has arrived.
 The simple knowledge
that positive energy has arrived however implies,
as we said, by fourier
transformation that any finite time interval has a nonvanishing
probability of being the time of measurement.
In general, the more precisely one measures the
arriving particle's energy
the more uncertain becomes the time of measurement
 and thus, whether the
time of measurement was in- or outside the light cone.
\mn
On the other hand, one may decide to measure instead
 the time of the
particles arrival very precisely. However, any
 sharp localisation
in time, like e.g. a gau{\ss}ian or even $\delta$-peak, always
fourier decomposes into positive and negative frequencies
 and must therefore
contain particle and antiparticle contributions:
The energy disturbance
through precise time measurement implies contributions of
pair creation. Even if it is only measured
whether the time of measurement is outside the
lightcone or not, the above
theorem in fourier theory tells, that all finite
 positive or negative
energy intervals have nonvanishing amplitudes, thus one could not
be certain whether positive energy, i.e. a particle, is found.
\mn
That there is no faster than light information
transport in quantum field theory has been pointed out
e.g. in \cite{fpl}. As we saw, the spacelike
propagation is merely a consequence
of the uncertainty principle, and in general no
quantum field theoretical
propagators could be restricted to the inside of the lightcone.
Independent of the actual dynamics and
 interactions, the propagators
must have this feature.
The deeper reason is,
that due to the energy-time commutation relation
the energy- and the time eigenbasis are related
 by fourier transformation
which implies the well known uncertainty relation.
\sn
However, and this is now a new aspect, any generalisation of
the Heisenberg algebra of 4-positions and 4-momenta
will of course
influence the propagator's behaviour through a modification
of the uncertainty principle. Let us
briefly discuss what can be expected
from the introduction of nonzero minimal uncertainties in
positions and momenta.
\mn
Nonzero minimal uncertainties in position
 and time measurements would
provide an absolutely lower limit
to the precision to which spacetime distances, and thus also light
cones could be described.  As we said, our uncertainty
relation for a minimal position uncertainty
has already appeared in the context of
string theory, which is not unexpected since this is
a theory of explicitly non-pointlike particles.
A minimal uncertainty in positions can also be motivated with
considerations of gravity in the ultraviolet.
The framework then expresses the old idea
of a minimal observable length. For the propagators this should
effectively imply an ultraviolet cutoff. We will here focus on the
regularisation aspect.
\mn
On the other hand, our motivation is also to understand
ordinary i.e. 'infrared' gravity, i.e. curvature
as a source of nonzero minimal uncertainties in
momentum and energy. It is easy to see, what it would mean
if e.g. time dependent curvature
implies nonzero minimal uncertainties in energy:
The sharp distinction between positive frequencies (i.e.
positive energies or particles
travelling forward in time) and negative
frequencies (i.e. negative energies or antiparticles travelling
backwards in time) would no longer be possible. The propagators
could reflect spontaneous
creation of particles and energy nonconservation.
On a general curved spacetime energy is of course
not conserved and also spontaneous particle creation from
gravity has been discussed in the literature, see
e.g. \cite{unruh,jarm}.
The concept of a particle as a irreducible representation
of the Poincar{\'e} algebra can then no longer be applied
straightforwardly. We will return to this question of
curvature and the external symmetry group in Section 4.
\mn
Here we are not yet trying to make this
contact to gravity explicit. Instead we will study how
nonzero minimal uncertainties
can be introduced into quantum field theory and whether this
has a regularising effect.
The basic idea of our approach is, to generalise the
Heisenberg algebra of 4-positions and 4-momenta which
also in quantum field theory 'sets the stage' of position
and momentum space.
Technically our proceeding is as follows:
\mn
While usually the path integral is performed over the
space of fields
where the fields are given in either the position
or the momentum eigenbasis,
we now rewrite the same path integral with the fields
given in the Bargmann Fock basis\footnote{Recall that
for the more general Heisenberg commutation relations
the representations alternative to the $x$ or $p$- representations
are of course not necessarily Bargmann Fock representations.}.
Once the formalism is rewritten in
the alternative basis we are independent
of the classical position or
momentum space. We do then no longer need or assume
the existence of 4-position or 4-momentum eigenvectors.
This allows us to generalise the Heisenberg algebra
of the $\x_{\mu}$ and $\p_{\nu}$. Analogously to section 2,
we can then abandon
the notion of points in position or momentum space.
Still all information on positions and momenta can be obtained from
the matrix elements of the symmetric position and momentum operators.
\mn
We will have to address the important problem of defining
the boundary conditions of the equations of motions
in the generalised framework.
The translation into Bargmann Fock space
of the split-up of fields in their
positive and negative frequency components, would not be
difficult. However we would also have to find the correct
generalisation of it for the noncommutative case.
As we said, we expect this to be related e.g. to the Unruh effect.
We postpone this discussion which would involve gravity. Let us
in the beginning work with the euclidean signature instead.

The generalisation of the commutation
relations and of the formalism itself
is then straightforward, also in four dimensions.
On the other hand, when generalising Lagrangians, we will
encounter the well known phenomenon that 'quantisation
removes degeneracy' (Dirac): Once former commutative variables become
noncommutative, there are then many different expressions
which would all reduce to the
same expression in the commutative case.
One needs to develop a guiding principle
 to determine the 'right' generalised
Lagrangians. We will only propose a working hypothesis.
With the example of a quadratically divergent loop
integration we then
demonstrate that the introduction of the nonzero minimal
uncertainties really has the power to regularise.

\subsection{Quantum field theory on Bargmann Fock space}
Using the path integral method we will now
study the properties of the generalised field theoretical
framework in the example of a charged scalar field with
$\phi^4$ self interaction, in four euclidean dimensions.
For the beginning we postpone the discussion of the noncommutative
generalisation of the Feynman boundary conditions
and work instead with the 4-dimensional
euclidean signature where the boundary conditions
are trivial\footnote{The reason being of course that
the Wick rotation changes the hyperbolic equation of motion into an
elliptic one, which then no longer has nontrivial
homogeneous solutions.}.

This standard theory we
reformulate, so that the fields then live on Bargmann Fock space.
Generalising Bargmann Fock space, we introduce nonzero minimal
uncertainties in positions and momenta. We study whether
this yields regularised behaviour of loop integrals. The
simple example given is the divergent first order correction
to the propagator.

\subsubsection{$\phi^4$-theory}
The generating functional for euclidean charged $\phi^4$ theory is
\be
Z[J,J^*]:= N \int D\phi\mbo D\phi^*\mbo e^{\int d^4x\mbo
\phi^* (\partial_i\partial_i - \mu^2)\phi
 - \frac{\lambda}{4!}(\phi \phi)^*\phi \phi + \phi^*J+J^*\phi}
\ee
where velocities and actions are measured as multiples
 of $c$ and $\hbar$.
\sn
We reintroduce the fundamental constants:
\be
Z[J,J^*] = N \int D\phi\mbo D\phi^* \mbo e^{1/\hbar \int d^4x\mbo
\phi^*(\partial_i\partial_i - \frac{M^2c^2}{\hbar^2})\phi
 - \frac{\lambda}{4!\hbar}(\phi\phi)^*\phi\phi + \phi^*J+J^*\phi}
\ee
The fields and the coupling constant
carry the units (e.g. in kg,m,s):
\be
[\phi] = \sqrt{\frac{kg}{s}},\qquad
[J]= \frac{1}{m^2}\sqrt{\frac{kg}{s}}, \qquad [\lambda]= 1
\ee
The theory is unchanged if we multiply the field by
a constant. Using the (arbitrary) positive
length $L$ we redefine: $\phi\rightarrow L \sqrt{\hbar} \phi$ :
\be
Z[J,J^*] = N \int D\phi\mbo D\phi^*\mbo e^{\int d^4x\mbo
\phi^* (-\frac{L^2}{\hbar^2}\p_i\p_i
- \frac{L^2M^2c^2}{\hbar^2})\phi
 - \frac{\lambda L^4}{4!}(\phi\phi)^*\phi\phi + \phi^*J+J^*\phi}
\ee
where we used $\p_j\phi(x) := -i\hbar\partial_j\phi(x)$.
The units are now:
\be [\phi] = 1/m^2,\qquad
[J]= 1/m^2, \qquad [\lambda]= 1
\ee
This formulation is
convenient for the transformation into the
Bargmann Fock space, because
as we will see it will keep the Bargmann Fock
 polynomials unitless.
\subsubsection{Fields on Bargmann Fock space}
Let us, term by term, transform the action functional
into Bargmann Fock space:
\mn
The source terms are simply the scalar product of two functions:
\be
\int d^4x\mbo \phi^*(x)J(x) = \overline{\phi(\etab)}\mbo
e^{\deta_i\detab_i}\mbo
J(\etab)\vert_0
\ee
\be
\int d^4x\mbo J^*(x)\phi(x) = \overline{J(\etab )}
e^{\deta_i\detab_i}\mbo
\phi(\etab)\vert_0
\ee
In the term quadratic in $\phi$ we recognise the scalar product
of the function $\phi$ with the
function $K.\phi$ where the operator $K$
is:
\be
K = (-L^2/\hbar^2)(\p_i\p_i+M^2c^2) \qquad
\mbox{ (summed over $i$)}
\ee
While $\p_j$ would in momentum space be a multiplication operator,
we know that it reads on Bargmann Fock space:
\be
\p_j.\phi(\etab) = \frac{i\hbar}{2L}(\etab_j -
\partial_{\etab_j})\phi(\etab)
\ee
so that $K$ acts on functions on Bargmann Fock space as:
\be
- \frac{L^2}{\hbar^2}(\p_i\p_i +M^2c^2).\phi(\etab)
= \left(\frac{1}{4}\sum_{i=1}^{4}(\etab_i-\detab_i)^2
- \frac{L^2M^2c^2}{\hbar^2}\right)\phi(\etab)
\ee
Thus, the quadratic term of the action reads:
$$
\int d^4x\mbo \phi^*(x) \left(- \frac{L^2}{\hbar^2}\p_i\p_i
- \frac{L^2M^2c^2}{\hbar^2}\right)\phi(x)
\qqq \qqq
$$
\be
\qquad \qquad \qquad \qqq
=  \overline{\phi(\etab )}\mbo e^{\deta_i\detab^i}
\left(\frac{1}{4}\sum_{i=1}^{4}(\etab_i-\detab_i)^2
- \frac{L^2M^2c^2}{\hbar^2}\right)\phi(\etab) \vert_0
\ee
The interaction term is the scalar product
of the function $\phi(x)\phi(x)$
with itself, where in $\phi(x)\phi(x)$ we recognise the pointwise
multiplication $*$.
\sn
Thus the interaction term reads in Bargmann Fock space:
\be
-\frac{\lambda L^4}{4!} \int d^4x\mbo
 \overline{\phi(x)\phi(x)}\phi(x)\phi(x)
=
-\frac{\lambda L^4}{4!} \overline{(\phi*\phi)(\etab )}
\mbo e^{\deta_i\detab^i}\mbo
(\phi *\phi)(\etab) \vert_0
\label{vertex}
\ee
where the pointwise multiplication $*$ is explicitly
(from Eq.\ref{star1dim}):
\be
(\phi *\phi)(\etab) = \frac{2}{3\pi L^2} \mbo
e^{\sum_{i=1}^{4}\left( \frac{1}{3}
 (\etab_i + \partial_{\etab^{\prime}_i}
+ \partial_{\etab^{\prime \prime}_i})^2 -\frac{1}{2} (\etab_i^2
+ \partial_{\etab^{\prime}_i}^2 +
 \partial_{\etab^{\prime \prime}_i}^2)
\right)} \phi(\etab^{\prime})\phi(\etab^{\prime\prime})\vert_0
\label{ptmul4dim}
\ee
Note that the scale $L$ drops out
when Eq.\ref{ptmul4dim} is put into Eq.\ref{vertex}.
\sn
The use of the weighted pointwise multiplication
$*^{\prime}$ instead
of $*$, would lead to a modified infrared behaviour. In
Bargmann Fock space it takes a simple form:
$$
-\frac{\lambda L^4}{4!} \int d^4x\mbo
\overline{\phi(x)\phi(x)}\phi(x)\phi(x)\mbo
e^{\frac{1}{2}\left(\frac{x}{L}\right)^2}
=
-\frac{\lambda L^4}{4!} \overline{(\phi*^{\prime}\phi)(\etab )}
\mbo e^{\deta_i\detab^i}\mbo
(\phi *^{\prime}\phi)(\etab) \vert_0
$$
\be
\qqq = \frac{-\lambda}{4\pi^2 4!}\mbo
\overline{\phi(\etab + \partial_{\etab^{\prime}})
\phi(\etab + \etab^{\prime})}
\mbo e^{\deta_i\detab^i}\mbo
\phi(\etab + \partial_{\etab^{\prime \prime}})
\phi(\etab + \etab^{\prime \prime})\vert_0
\ee
Since $L$ can be chosen arbitrarily large, we could,
up to the far infrared behaviour, approximate $*$
by the much easier to handle $*^{\prime}$.
\sn
The transformation into Bargmann Fock space is, like
the transformation into e.g. momentum space, just a linear change
of basis in the vector space of functions. The determinant
of its Jacobian is a constant and can be absorbed in the
overall factor of the path integral. We do thus not introduce
anomalies.
\subsubsection{Feynman rules}
When working in the position or the momentum
eigenbasis of the function space, we are formally dealing with
an uncountably infinite number of basis states, namely the
set of $\delta$- functions or the set of plane waves. However,
the function space is actually separable. This means that
there are bases
of only a countable infinite number of vectors, which do
also span the function space. The
Bargmann Fock polynomials $\etab^m$,
which in position space are the Hermite functions, form such a
countable basis, which spans the Hilbert space of square
integrable functions (and
in which we can expand even the
non-normalisable $\delta$- functions and
plane waves, see Chapter 2).

Thus, we will now expand our $\phi$
in the discrete basis of Bargmann Fock
 polynomials $\etab^m$. We will
eventually express the action functional in
 terms of the coefficients
that the vectors (i.e. the functions) and the operators have
in this orthonormal basis:
\be
\phi(\etab) = \sum_{r_1,r_2,r_3,r_4 =0}^{\infty}
\phi_{\vec{r}}\mbo \frac{\etab_1^{r_1}
 \etab_2^{r_2} \etab_3^{r_3}
\etab_4^{r_4}}{\sqrt{r_1! r_2! r_3! r_4!}}
\ee
\be
J(\etab) = \sum_{r_1,r_2,r_3,r_4 =0}^{\infty}
J_{\vec{r}}\mbo \frac{\etab_1^{r_1}
\etab_2^{r_2} \etab_3^{r_3}
\etab_4^{r_4}}{\sqrt{r_1! r_2! r_3! r_4!}}
\ee
Thus a function $\phi$ is now represented by its coefficient vector
$\phi_{\vec{r}} = \phi_{r_1,r_2,r_3,r_4}$ with the
indices $r_i = 0,1,2,...\infty$, ($i=1,...,4$).
We ordered the polynomials because, due to commutativity, e.g.
\be
 \frac{\etab_1^{r_1} \etab_2^{r_2} \etab_3^{r_3}
\etab_4^{r_4}}{\sqrt{r_1! r_2! r_3! r_4!}}
=
 \frac{\etab_4^{r_4} \etab_2^{r_2} \etab_3^{r_3}
\etab_1^{r_1}}{\sqrt{r_1! r_2! r_3! r_4!}}
\ee
so that the set of all (nonordered) polynomials would have been
overcomplete. In our generalised
formalism, when the polynomials in $\etab$'s become
noncommutative, the ordered polynomials will still
form a basis\footnote{This requirement, i.e. technically the
invariance of the Poincar{\'e} series (i.e. of the dimensionalities
of the subspaces of polynomials of equal grade),
was one of the key
conditions in the derivation of the generalised commutation
relations, see \cite{ak-lmp-bf}.}.
\sn
The coefficient matrix of the quadratic operator
is then calculated as
\be
M_{\vec{r}\vec{s}} =
\frac{\overline{\etab_1^{r_1} \etab_2^{r_2} \etab_3^{r_3}
\etab_4^{r_4}}}{\sqrt{r_1! r_2! r_3! r_4!}}
\mbo e^{\deta_i\detab^i}\mbo
\left( -\frac{1}{4}\sum_{i=1}^{4}(\etab_i - \partial_{\etab_i})^2
+ \frac{L^2M^2c^2}{\hbar^2}\right)
\frac{\etab_1^{s_1} \etab_2^{s_2} \etab_3^{s_3}
\etab_4^{s_4}}{\sqrt{s_1! s_2! s_3! s_4!}}\mbo \vert_0
\ee
and the interaction term has the coefficients:
\be
V_{\vec{t}\vec{u}\vec{v}\vec{w}} =
\frac{\overline{\etab_1^{t_1}
 \etab_2^{t_2} \etab_3^{t_3} \etab_4^{t_4}
* \etab_1^{t_1} \etab_2^{t_2} \etab_3^{t_3} \etab_4^{t_4}}
}{\sqrt{t_1! t_2! t_3! t_4! u_1! u_2! u_3! u_4!}}
\mbo e^{\deta_i\detab_i} \mbo
\frac{{\etab_1^{v_1} \etab_2^{v_2} \etab_3^{v_3} \etab_4^{v_4}
* \etab_1^{w_1} \etab_2^{w_2} \etab_3^{w_3} \etab_4^{w_4}}
}{\sqrt{v_1! v_2! v_3! v_4! w_1! w_2! w_3! w_4!}}\mbo \vert_0
\ee
Note that the path integration can be written as the
product of a \it countably \rm infinite number of integrations:
\begin{eqnarray}
\int D\phi\mbo D\phi^* \mbo e^{-S[\phi(x),\phi^*(x)]}
&=& N \int D\phi\mbo D\overline{\phi}
e^{-S[\phi(\etab),\overline{\phi(\etab)}]}\\
 &=& N \int \prod_{r_1,r_2,r_3,r_4=0}^{\infty}
d\phi_{r_1,r_2,r_3,r_4}
\mbo d\phi^*_{r_1,r_2,r_3,r_4}
\mbo e^{-S[\phi_{\vec{r}},\phi^*_{\vec{r}}]}
\nonumber
\end{eqnarray}
The Feynman rules are derived in the standard way.
The generating functional now reads:
\be
Z[J,J^*] = N \int D\phi \mbo D\phi^* \mbo e^{-\mbo
\phi_{\vec{r}}^*\mbo M_{\vec{r}\vec{s}}
\mbo \phi_{\vec{s}}\mbo +\mbo \phi_{\vec{r}}^* J_{\vec{r}}
 \mbo+\mbo
J^*_{\vec{r}} \phi_{\vec{r}}\mbo
-\frac{\lambda L^4}{4!}
V_{\vec{t}\vec{u}\vec{v}\vec{w}}
\phi_{\vec{t}}^* \phi_{\vec{u}}^* \phi_{\vec{v}} \phi_{\vec{w}}}
\ee
Here, each index vector denotes four indices, corresponding
to the four
euclidean dimensions, e.g.
$\vec{r} = r_1,r_2,r_3,r_4$ where each index is summed over, e.g.
$ r_2 = 0,1,2,...\infty$. Pulling the interaction term
 in front of the
integral yields:
\be
Z[J,J^*] = N e^{-\frac{\lambda L^4}{4!}
V_{\vec{r}\vec{s}\vec{t}\vec{u}}
\frac{\partial}{\partial J_{\vec{r}}}
\frac{\partial}{\partial J_{\vec{s}}}
\frac{\partial}{\partial J_{\vec{t}}}
\frac{\partial}{\partial J_{\vec{u}}}}
\int D\phi\mbo D\phi^*
\mbo e^{-\mbo \phi_{\vec{r}}^*\mbo M_{\vec{r}\vec{s}}
\mbo \phi_{\vec{s}}\mbo +\mbo \phi_{\vec{r}}^* J_{\vec{r}} \mbo+\mbo
J^*_{\vec{r}} \phi_{\vec{r}}}
\ee
We rearrange the remaining integrand:
\be
Z[J,J^*] = N e^{-\frac{\lambda L^4}{4!}
V_{\vec{r}\vec{s}\vec{t}\vec{u}}
\frac{\partial}{\partial J_{\vec{r}}}
\frac{\partial}{\partial J_{\vec{s}}}
\frac{\partial}{\partial J_{\vec{t}}}
\frac{\partial}{\partial J_{\vec{u}}}}
\int D\phi D\phi^*
 \mbo e^{-\mbo (\phi_{\vec{r}}^* -  J^*_{\vec{s}}
M^{-1}_{\vec{s}\vec{r}}) M_{\vec{r},\vec{t}}
(\phi_{\vec{t}} - M^{-1}_{\vec{t}\vec{u}} J_{\vec{u}})\mbo
+\mbo J^*_{\vec{r}} M_{\vec{r} \vec{s}} J_{\vec{s}}}
\ee
The path integral can now be absorbed in the overall constant:
\be
Z[J,J^*] = N^{\prime} e^{-\frac{\lambda L^4}{4!}
V_{\vec{r}\vec{s}\vec{t}\vec{u}}
\frac{\partial}{\partial J_{\vec{r}}}
\frac{\partial}{\partial J_{\vec{s}}}
\frac{\partial}{\partial J_{\vec{t}}}
\frac{\partial}{\partial J_{\vec{u}}}}
\mbo e^{J^*_{\vec{r}} M_{\vec{r} \vec{s}} J_{\vec{s}}}
\ee
For the Feynman rules we read off:
\be
\mbox{free propagator}\mbox{:\quad }
\Delta_o(\vec{a},\vec{b}) = M^{-1}_{\vec{a}\vec{b}}
\ee
\be
\mbo \mbox{vertex: \qquad }
\Gamma_0(\vec{a},\vec{b},\vec{c},\vec{d}) =
-\frac{\lambda L^4}{4!} V_{\vec{a}\vec{b}\vec{c}\vec{d}}
\ee
\subsubsection{Loop integration}
The usual four dimensional loop integrations,
e.g. over momentum space, now become four discrete summations.
\sn
For example, differentiating $Z[J,J^*]$ with respect
to $J^*_{\vec{a}}$ and $J_{\vec{b}}$ at vanishing sources,
and neglecting the disconnected terms, yields for the
first\footnote{Recall that one could formally avoid
this tadpole diagram by, in the operator formalism, normal
ordering
the interaction lagrangian. This would amount
 to an (infinite) mass
renormalisation.}
order correction to the propagator:
\be
\Delta(\vec{a},\vec{b}) = M^{-1}_{\vec{a}\vec{b}}
\quad
- 4 \frac{\lambda L^4}{4!}  \cdot
\sum_{r_1,r_2,r_3,r_4 = 0}^{\infty}\mbo
\sum_{s_1,s_2,s_3,s_4 = 0}^{\infty}
V_{\vec{a}\vec{r}\vec{s}\vec{b}}
M^{-1}_{\vec{a}\vec{r}}
M^{-1}_{\vec{r}\vec{s}}
M^{-1}_{\vec{s}\vec{b}} + ...
\label{tad}
\ee
The correction is the tadpole term which in
momentum space reads, up to the external legs and a constant,
\be
-\frac{\lambda}{6} \int d^4p\mbo \frac{1}{p_i p^i + M^2c^2} =
\mbox{ divergent}
\label{tadpolemom}
\ee
and which is a quadratic divergency. If we now calculated
$M^{-1}$ and $V$ explicitly and applied them in Eq.\ref{tad}
we could only
recover that the summations are divergent. Let us therefore turn
to the generalised framework where we will
work with the explicit matrices $M^{-1}$ and $V$ and find that
this graph becomes regularised.

\subsection{Generalised quantum field theory on Bargmann Fock space}
Our strategy in generalising the quantum field theoretical
framework was to first reformulate the path integral so that
the fields live on Bargmann Fock, rather than
on position or momentum space. We now generalise the
commutation relations of the position and momentum operators,
completely like in section 2, only in four dimensions now.
In this way we
introduce nonzero minimal uncertainties in positions and momenta,
and abandon position or momentum eigenfunctions.
We can use the techniques developed in
section 2 to transform the fields, operators and
the scalar product into noncommutative Bargmann Fock space.

Thus the kinematics, i.e. the framework itself, can be generalised
straightforwardly. However, the generalisation of the
dynamics i.e. the generalisation of Lagrangians is not
straightforward.
\mn
There are at least three different sources of ambiguity
 in generalising
Lagrangians:
\mn
\bf 1. \rm A generalised Lagrangian need not necessarily
become the old Lagrangian as we remove the generalisation
parameter (i.e. here as $q\rightarrow 1$). In this limit we need
to recover the old theory only where the old theory is
actually valid. This
means, that the generalised Lagrangian may deviate
in the far infrared or in the far ultraviolet,
even as $q\rightarrow 1$. Ultimately
we would hope that the correctly generalised Lagrangians allow
for the description of the gravity effects that appear
at these scales.
\sn
\bf 2. \rm We need a generalised notion of pointwise multiplication.
In order to describe local interactions,
usual quantum field theoretical Lagrangians
rely on the notion of pointwise multiplication $*$ of fields.
Usually we can
transform fields on Bargmann Fock space into position space,
multiply them pointwise and transform the result back into
Bargmann Fock space. In this way one obtains the
pointwise multiplication of Bargmann Fock functions.
However, in the generalised case there is no transformation
into a position space any more because we
abandon position eigenstates i.e.
we abandon the very notion of a point in position
space. Thus,
in the generalised case the 'pointwise'
multiplication of functions on Bargmann Fock space, can be
defined quite arbitrarily. As one removes the generalisation
parameter it has to reduce to
ordinary pointwise multiplication but, as we said in 1,
possibly only up to the far UV or IR.
\sn
\bf 3. \rm 'Quantisation removes degeneracy' (Dirac) originally
meant effects that appear as the phase space becomes noncommutative
i.e. that appeared by quantisation.
How would e.g. a function $f(x,p)$ (with
$[x,p]=0$) read when the variables $x,p$ become
noncommutative: $[x,p]=i\hbar$. There is then obviously no unique
$f(x,p)$ due to
ordering ambiguities. An analogous ambiguity appears here, as
we generalise the commutation relations even further, so that also
the position operators among themselves,
as well as the momentum operators among themselves
become noncommutative.
\bn
A guiding principle is needed that determines the
'correct' generalisation of
La\-grang\-ians. This guiding principle must of course
be found on a physical basis. Ultimately, a 'correctly' generalised
Lagrangian should allow to understand the generalisation parameters,
more or less directly, as gravitational degrees of freedom.
Large scale gravity i.e. general relativity is well confirmed.
It must of course be studied whether in our approach one can at least
reproduce this
large scale gravity as an effective theory in the medium and large
distances. Work in this direction is in progress.
On the other hand, only little is known
about gravity in the ultraviolet,
i.e. at small distances like the Planck length.
At this stage we can only speculate whether e.g.
positions and momenta then
play with reversed r{\^o}les, i.e. that then
 noncommutativity of the
position operators becomes crucial (and may even
reflect a 'curvature-like'
feature of momentum space). The beautyful duality
of the positions and
momenta is usually thought to be restricted to the kinematics.
The main reason is of course that interactions appear to be
local only in position space, a fact which singles
 out position space.
However, as we said, the very concept of locality
needs to be reconsidered for a quantum theory that
 includes gravity.
Our formalism offers a framework
in which such new ideas can be formulated and studied.
\mn
We have not yet developed representations of our fully generalised
commutation relations \ref{genHi}-\ref{genHf}. We can
therefore not yet work with the correspondingly generalised
quantum field theoretical framework. However, we do have
representations of two sub classes of our generalised
commutation relations. We will study the generalisation of the
quantum field theoretical framework for these examples.
Once representations for the fully general commutation relations
are found, it should be straightforward to also work out the
correspondingly generalised quantum field theoretical framework.
\mn
As we said, a guiding principle for the generalisation
of arbitrary Lagrangians is needed.
We do yet attempt to propose such a
guiding principle. Instead we will deal only with a working
hypothesis for the generalisation of the Lagrangian of the
bosonic scalar charged particle with self
interaction. We focus on the technical study whether
ultraviolet divergencies can become regularised through
nonzero minimal uncertainties.
\subsubsection{Generalised $\phi^4$ theory}
We study features of the generalised quantum field
theoretical framework
in the simple example of the $\phi^4$ theory.
The formulas for our first sub class of generalised
commutation relations will be given explicitly. From these,
the corresponding
expressions for the case of our second sub class of commutation
relations can easily be derived.
In particular, our main result, namely the convergence of our
example graph, will be derived for both cases.
\sn
The original generating functional, with the fields defined
on position space, reads:
\be
Z[J,J^*] = N \int D\phi\mbo D\phi^*\mbo e^{\int d^4x\mbo
\phi^* (-\frac{L^2}{\hbar^2}p_ip_i
- \frac{L^2M^2c^2}{\hbar^2})\phi
 - \frac{\lambda L^4}{4!}(\phi\phi)^*\phi\phi + \phi^*J+J^*\phi}
\label{ordlagr}
\ee
The generalised Lagrangian need not to
reduce exactly to the above Lagrangian in the
limit $q\rightarrow 1$.
As we said, we would allow the generalised
Lagrangian to deviate as $q\rightarrow 1$
from Eq.\ref{ordlagr} in the IR and UV,
since the ordinary theory does not describe the gravity effects
in the large and in the very small distances.
\sn
Let us propose the following working hypothesis:

We will use the freedom in the generalisation to
work with a generalised Lagrangian that, as
$q\rightarrow 1$, reduces to the ordinary
$\phi^4$ Lagrangian, but deviates in the infrared
behaviour. We introduce a modified long distance
behaviour of the
quadratic term, which is mathematically quite natural
and one could even speculate about a possible
physical justification. Our modification of
the infrared behaviour of also the interaction term
has its reason
only in that this modified interaction term is more
straightforward to generalise and easier to handle than the
original interaction term.

We do not claim that this is physically sensible.
Our main objective
here, is to study our generalised quantum
 field theoretical framework
with the example of a toy Lagrangian that is mathematically
easy to deal with.
\mn
Explicitly our working hypothesis for the generalised $\phi^4$
theory is:
\begin{eqnarray}
Z[J,J^*] = N\int D\phi\mbo D\phi^*& &
e^{\overline{\phi(\etab)}\mbo
e_{1/q}^{\partial_{\eta_i}\partial_{\etab_i}}
\left( - \etab_i \partial_{\etab_i} - 2
- \frac{L^2M^2c^2}{\hbar^2}\right)\phi(\etab)\mbo \vert_0}
\nonumber\\
&{\cdot}& e^{-\frac{\lambda}{4\pi^2 4!} \overline{\phi(\etab +
\partial_{\etab^{\prime}}) \phi(\etab + \etab^{\prime})}\mbo
e_{1/q}^{\partial_{\eta_i}\partial_{\etab_i}}\mbo
\phi(\etab + \partial_{\etab^{\prime \prime}})
\phi(\etab + \etab^{\prime \prime})\mbo \vert_0}
\label{genlagr} \\
&{\cdot}& e^{\overline{\phi(\etab)}\mbo
e_{1/q}^{\partial_{\eta_i}\partial_{\etab_i}}\mbo
J(\etab)\mbo \vert_0\mbo + \mbo \overline{J(\etab)}\mbo
e_{1/q}^{\partial_{\eta_i}\partial_{\etab_i}}\mbo
\phi(\etab)\mbo \vert_0}\nonumber
\end{eqnarray}
It is clear that our
Lagrangian reduces, for $q\rightarrow 1$, and then
transformed into position space, to the Lagrangian:
\be
{\cal{L}}^{\prime} = \phi^* (-\frac{L^2}{\hbar^2}\p_i\p_i
- \frac{\x_i\x_i}{4 L^2}
- \frac{L^2M^2c^2}{\hbar^2})\phi
- \frac{\lambda L^4}{4!}(\phi\phi)^*\phi\phi \mbo
e^{\frac{x_ix_i}{2 L^2}}
+ \phi^*J+J^*\phi
\ee
Our change in the infrared behaviour of the
 interaction term is, to work
with the pointwise multiplication $*^{\prime}$ instead of $*$.
The change of the long distance behaviour in the kinetic term
is the addition of the term $\x_i\x_i/4L^2$.
We thus work with the operator $\etab_i\detab_i$ which is
diagonal in the Bargmann Fock basis, so that
also the free propagator will be diagonal.
Physically, we are
treating the free particle as a particle in a harmonic
oscillator potential with the backdriving force negligibly
small up to large distances (we assume $L$ large).
This of course means that we are breaking the 10- dimensional
Poincar{\'e} symmetry down to the 6- dimensional
Lorentz symmetry. However, the Lorentz
symmetry $SO(3,1)$ is actually
enlarged to the unitary
group\footnote{The $SU(3,1)$ has a quantum group generalisation
$SU_q(3,1)$, see \cite{FRT}.}
$SU(3,1)$ which has 15, i.e. even more
dimensions than the Poincar{\'e} group.
On the other hand, the appearance of a harmonic
oscillator potential from gravity seems not very likely. Note
however that e.g. a test particle inside a homogeneous,
spherical symmetric gravitating energy distribution
experiences gravity as a harmonic oscillator potential. (This is
e.g. how a particle inside the earth
 experiences the earth's gravity.)

Let us however stress that these arguments only apply to
the toy model that we are considering here. The correct
generalisation of the dynamics and of the
Poincar{\'e} algebra will go beyond the
quantum group symmetric case and can of course only
be obtained from a detailed analysis
of the physical origin of the
modifications to the commutation relations. We will begin
this discussion in Section 4.
\mn
Our working hypothesis for the generalised Lagrangian
reduces, as $q\rightarrow 1$ to the ordinary $\phi^4$ Lagrangian
for short distances, i.e. in the ultraviolet.
We will see in an example that our
generalisation of the framework itself, i.e.
of the commutation relations
and thus the introduction of nonzero minimal
uncertainties in positions
and momenta, does affect e.g. the ultraviolet
in a regularising way.

\subsubsection{Fields on Bargmann Fock space}
The path integral in Eq.\ref{genlagr} is
straightforward to evaluate,
whether $q=1$ or $q > 1$: Summing over all
 fields means to sum over
all Bargmann Fock functions, i.e. to sum over all distinct
polynomials (or power series) in the $\etab$'s.
Each such Bargmann Fock
function is to be put into the exponentiated
action functional in order to obtain its amplitude.
For clearity let us consider simple examples:
\sn
The simplest Bargmann Fock function is
 the constant $\phi(\etab) :=1$.
We calculate the amplitude i.e. the contribution
to the path integral Eq.\ref{genlagr} of
$\phi(\etab)=1=\overline{\phi(\etab)}$. From the source terms
we get:
\begin{eqnarray}
 & &e^{\mbo 1\mbo
e_{1/q}^{\partial_{\eta_i}\partial_{\etab_i}}\mbo J(\etab)
\mbo \vert_0
\mbo + \mbo \overline{J(\etab)}
e_{1/q}^{\partial_{\eta_i}\partial_{\etab_i}}
\mbo 1\mbo \vert_0}
\nonumber\\
 &=&e^{\mbo 1\mbo
\sum_{r=0}^{\infty}\frac{
 {\partial_{\eta_i}\partial_{\etab_i}} }{[r]!}
\mbo J(\etab)\mbo \vert_{\eta = 0, \etab = 0}
\mbo + \mbo \overline{J(\etab)}
\sum_{r=0}^{\infty}\frac{{\partial_{\eta_i}\partial_{\etab_i}}}{[r]!}
\mbo 1\mbo \vert_{\eta=0,\etab=0}}
\nonumber\\
 &=&e^{\mbo J_{0,0,0,0}\mbo + \mbo J_{0,0,0,0}^*}
\nonumber
\end{eqnarray}
where we expanded the source in the basis of the
ordered orthonormal polynomials:
\be
J(\etab)=\sum_{s_1,s_2,s_3,s_4=0}^{\infty}
J_{s_1s_2s_3s_4}\mbo
\frac{\etab_1^{s_1}\etab_2^{s_2}\etab_3^{s_3}\etab_4^{s_4}}{
\sqrt{[s_1]![s_2]![s_3]![s_4]!}}
\ee
This expansion of an arbitrary $J$ is possible because
even now that the $\etab$'s are noncommutative, we can still rewrite
each polynomial in the ordered form, see \cite{ak-lmp-bf}.
Recall that in general the evaluation procedure is to first commute all
$\deta$'s to the left and all $\detab$'s to the right, using the
(generalised) Leibniz rules and commutation relations
among the $\etab$'s, among the $\detab$'s and their complex conjugates.
The evaluation is to be carried out at $\eta =0$ and $\etab=0$ i.e.
we keep only the constant terms and set all other terms equal zero.
\sn
In the term quadratic in $\phi$ and in the interaction term the
derivatives of the $q$-exponential do now not find $\etab$'s
to act on, so that we can immediately evaluate.
Thus the amplitude for $\phi(\etab) =1, \overline{\phi(\etab)}=1$ is:
\be
\mbox{ampl}[\phi(\etab)=1,\overline{\phi(\etab)}=1,J,J^*] = e^{
\mbo -2 \mbo -\mbo \frac{L^2M^2c^2}{\hbar^2}\mbo - \mbo
\frac{\lambda}{4\pi^24!} \mbo
+ \mbo J_{0,0,0,0}\mbo + \mbo J_{0,0,0,0}^*}
\ee
In the calculation of the amplitude for e.g.
$\phi(\etab):=\etab_3, \overline{\phi(\etab)}=\eta_3$
there are now some differentiations to be evaluated. E.g. the
contribution of the interaction term is:
\begin{eqnarray*}
 & &-\frac{\lambda}{4\pi^24!}\mbo
\overline{(\etab_3 + \partial_{\etab_3^{\prime}})(\etab_3 +
\etab^{\prime}_3)}\mbo
e_{1/q}^{\partial_{\eta_i}\partial_{\etab_i}}\mbo
(\etab_3 + \partial_{\etab_3^{\prime}})(\etab_3 +
\etab^{\prime}_3)\mbo \vert_0\\
 &=&-\frac{\lambda}{4\pi^24!}\mbo
(\eta^2_3 + 1)\mbo
e_{1/q}^{\partial_{\eta_i}\partial_{\etab_i}}\mbo
(\etab^2_3 + 1)\mbo \vert_0\\
 &=& -\frac{\lambda}{4\pi^24!}\mbo ([2] + 1) =
-\frac{\lambda}{4\pi^24!} (q^2+2)
\end{eqnarray*}
The amplitude is then easily found to be:
\be
\mbox{ampl}[\phi(\etab)=\etab_3,\overline{\phi(\etab)}=\eta_3,J,J^*]
= e^{\mbo -1\mbo -\mbo 2\mbo -
 \mbo \frac{L^2M^2c^2}{\hbar^2}\mbo
-\mbo \frac{\lambda}{4\pi^24!} (q^2+2)\mbo
+\mbo J_{0,0,1,0}\mbo + \mbo J^*_{0,0,1,0}}
\ee
In this way the amplitudes for arbitrary $\phi,\overline{\phi}$
can be calculated: In the
action functional, using the commutation
relations, the $\partial_{\eta}$'s and
$\partial_{\etab}$'s are commuted to the left and right.
We evaluate at $0$ i.e. we keep
only the constants. This yields a number, the exponential of which is
the amplitude that corresponds to the field considered.
\mn
The integrations over position or momentum space have
become integrations over Bargmann Fock space, which can be
written in terms of algebraic differentiations and summations.
The $n$- point functions $\Gamma$ now have
their arguments in Bargmann Fock
space rather than in position or momentum space.
Nevertheless all information on positions and momenta
can still be obtained.
\mn
In the ordinary framework the unitary
transformations given in section 2 could be used
to transform the propagators,
vertices, $n$-point functions or $S$-matrix
elements into momentum or position space.
Let us consider e.g.
a field given as a Bargmann Fock function $\phi(\etab)$.
In order to obtain its component $\phi(p^{\prime})$ of
a specific 4-momentum $p^{\prime}$ one simply calculates
(in Bargmann Fock space)
the scalar product of the function $\phi(\etab)$ with
the momentum eigenvector $\phi_{p^{\prime}}(\etab)$.
The form of the momentum eigenvectors in Bargmann Fock
space can either be calculated directly in Bargmann Fock space
or can also easily be obtained from its form
in the position or in the momentum eigenbasis,
using e.g. Eq.\ref{mombf}.
In the generalised
framework there are no more momentum (or position) eigenvectors
in the representation of the Heisenberg algebras.
We can however still calculate the scalar product with e.g. those
Bargmann Fock polynomials which describe particles that are
relatively localised in momentum (or position) space, but
of course not more precisely than $\Delta p_0$ or $\Delta x_0$.
The calculation of these approximating polynomials
was discussed in section 2.2.3 .

\subsubsection{Feynman rules}
Like in Sec. 3.1.3 it is more convenient for the derivation
of the Feynman rules
to work simply with the coefficients of the fields
 in the orthonormal basis
of ordered polynomials:
\be
\phi(\etab)=\sum_{s_1,s_2,s_3,s_4=0}^{\infty}
\phi_{s_1s_2s_3s_4}\mbo
\frac{\etab_1^{s_1}\etab_2^{s_2}\etab_3^{s_3}\etab_4^{s_4}}{
\sqrt{[s_1]![s_2]![s_3]![s_4]!}}
\ee
In this basis the generating functional has the form:
\be
Z[J,J^*] = N \int D\phi \mbo D\phi^* \mbo e^{-\mbo
\phi_{\vec{r}}^*\mbo M_{\vec{r}\vec{s}}
\mbo \phi_{\vec{s}}\mbo +\mbo \phi_{\vec{r}}^* J_{\vec{r}}
 \mbo+\mbo
J^*_{\vec{r}} \phi_{\vec{r}}\mbo
- \frac{\lambda}{4\pi^2 4!}
\mbo V_{\vec{t}\vec{u}\vec{v}\vec{w}}
\phi_{\vec{t}}^* \phi_{\vec{u}}^* \phi_{\vec{v}} \phi_{\vec{w}}}
\ee
The term quadratic in $\phi$, i.e. the matrix $M$ is actually
diagonal:
\be
M_{\vec{r}\vec{s}} = \left([r_1 + r_2 + r_3 + r_4] + 2
+ \frac{L^2M^2c^2}{\hbar^2}\right) \delta_{\vec{r},\vec{s}}
\ee
The usual derivation of the Feynman rules goes through and we
obtain the free propagator:
\be
\Delta_0(\vec{r},\vec{s}) =
\frac{1}{[r_1 + r_2 + r_3 + r_4] + 2
+ \frac{L^2M^2c^2}{\hbar^2}}\mbo \delta_{\vec{r},\vec{s}}
\ee
We calculate the simplest matrix elements of the vertex:
$$
\Gamma_0(\vec{0},\vec{r},\vec{0},\vec{s}) =
\Gamma_0(\vec{r},\vec{0},\vec{0},\vec{s}) =
\Gamma_0(\vec{0},\vec{r},\vec{s},\vec{0}) =
\Gamma_0(\vec{r},\vec{0},\vec{s},\vec{0}) =
$$
\be
= - \frac{\lambda}{4\pi^2 4!}
V_{\vec{0}\vec{r}\vec{0}\vec{s}} = ... =
- \frac{\lambda}{4\pi^2 4!}\mbo \delta_{\vec{r},\vec{s}}
\ee
Clearly all other matrix elements of the vertex are also finite.
\subsubsection{Loop integration}
Let us now study the example of the first order correction to the
propagator, namely the truncated tadpole graph. In the usual,
unregularised theory it is quadratically
divergent, see Eq.\ref{tadpolemom}.
We calculate the off shell amplitude $\Sigma(\vec{0},\vec{0})$ for
an in- and an out-going Gau{\ss}ian wavepacket
 $\phi(\etab) =1$. From the
usual perturbative expansion
\be
\Delta(\vec{a},\vec{b}) = M^{-1}_{\vec{a}\vec{b}}
\quad
- \frac{\lambda}{\pi^2 4!} \cdot
\sum_{r_1,r_2,r_3,r_4 = 0}^{\infty}\mbo
\sum_{s_1,s_2,s_3,s_4 = 0}^{\infty}
V_{\vec{a}\vec{r}\vec{s}\vec{b}}
M^{-1}_{\vec{a}\vec{r}}
M^{-1}_{\vec{r}\vec{s}}
M^{-1}_{\vec{s}\vec{b}} + ...
\ee
this amplitude is:
\begin{eqnarray}
\Sigma(\vec{0},\vec{0}) &=& - \frac{\lambda}{4\pi^2 4!} \cdot
\sum_{r_1,r_2,r_3,r_4 = 0}^{\infty}\mbo
\sum_{s_1,s_2,s_3,s_4 = 0}^{\infty}
V_{\vec{0}\vec{r}\vec{s}\vec{0}}
M^{-1}_{\vec{r}\vec{s}} \\
 &=& - \frac{\lambda}{\pi^2 4!} \sum_{r_1,r_2,r_3,r_4 = 0}^{\infty}
\frac{1}{[r_1 + r_2 + r_3 + r_4] + 2
+ \frac{L^2M^2c^2}{\hbar^2}} \nonumber\\
 &=& - \frac{\lambda}{\pi^2 4!} \sum_{r_1,r_2,r_3,r_4 = 0}^{\infty}
\frac{1}{\frac{q^{2(r_1 + r_2 + r_3 + r_4)}-1}{q^2-1} + 2
+ \frac{L^2M^2c^2}{\hbar^2}} \nonumber
\end{eqnarray}
This is actually finite, i.e.
 $\vert \Sigma(\vec{0},\vec{0})\vert < \infty$,
because for large $r_i$ the sum goes like
$$
(q^2-1) \sum_{r_1,r_2,r_3,r_4 = 0}^{\infty}
q^{-2(r_1 + r_2 + r_3 + r_4)}
$$
which is the product of four geometrical series:
$$
= (q^2-1) \left(\frac{1}{1-q^{-2}}\right)^4
$$
Thus, the quadratic ultraviolet divergency
has become a regularised function in the
generalisation parameter $q$. It has, of course, a pole at $q=1$.
\sn
Since we had not explicitly
modified the ultraviolet behaviour of the Lagrangian, we
interpret this result such that this ultraviolet
regularisation is built into the generalised
framework itself.
\mn
Let us check whether our second sub class of generalised
Heisenberg commutation relations also leads to the
regularisation of this graph. It is not difficult to see
that the construction of the generalised $\phi^4$
theory goes through straightforwardly. We now have to check the
convergence of the loop summation:
\be
\sum_{r_1,r_2,r_3,r_4=0}^{\infty}
\frac{1}{[r_1]_{q_1}+[r_2]_{q_2}+[r_3]_{q_3}+[r_4]_{q_4}+
2+ \frac{L^2M^2c^2}{\hbar^2}}
\ee
To see that this is convergent, note first that it can be
majorised by
\be
\sum_{r_1,r_2,r_3,r_4=0}^{\infty}
\frac{1}{[r_1]_{q^{\prime}}+[r_2]_{q^{\prime}}+
[r_3]_{q^{\prime}}+[r_4]_{q^{\prime}}+
2+ \frac{L^2M^2c^2}{\hbar^2}}
\ee
where $q^{\prime}$ is the smallest of the $q_i$ (recall that
all $q_i>1$). We now use the 'rotational' symmetry in the discrete
summation space ${\bf N\rm}^4$ of the $r_i$ to
majorise this multiple sum
by a simpler sum of the form
\be
\sum_{r=0}^{\infty} V(r) \frac{1}{q^r+ \mbox{const}}
\label{fsu}
\ee
Here $V(r)$ is number of terms that are to be summed over
in the 'layer' determined by $r^2\le \sum_{i=1}^{4}r_i^2
\le (r+1)^2$, which is essentially proportional to $r^3$.
Thus, since the denominator in the sum Eq.\ref{fsu} grows
exponentially, a simpe ratio test proves the convergence
of our sum.
\mn
We do not have a proof that all the generalised
commutation relations that imply nonzero minimal
uncertainties lead in general to the
convergence of loop graphs. The above results
however appear to support the intuitive expectation that
the introduction of nonzero minimal uncertainties does lead
to effective cutoffs.

\section{On the Poincar{\'e} algebra and the origins
of minimal uncertainties}
The Poincar{\'e} group plays
a fundamental r{\^o}le for the particle concept
in quantum field theory.
A detailed study of the external 'symmetry' will therefor be
needed for our generalised framework
in order to properly define spin and statistics in the
generalised setting, and also in order to check fundamental
properties like microcausality and unitarity.
We can not yet answer these important questions here,
let us however indicate a
possible start for these studies.
\sn
We note first that if a generalised Poincar{\'e} algebra,
generated by rotations and translations can be defined
in the generalised case, it should still be contained in the
Heisenberg algebra. We will therefore first have to
clarify if and how a physical origin of the generalisation
Eqs.\ref{genHi}-\ref{genHf}
of the commutation relations of the Heisenberg algebra can
be made precise.

\subsection{Ultraviolet
generalisation of the uncertainty relations}
Let us mention that the way in which nonzero minimal
uncertainties in positions enter our framework
appears to be in agreement with
results of quantum gravity and string theory,
see \cite{gsw,amati,konishi,maggiore}, where
effective uncertainty relations have been derived, which
are very similar to our uncertainty
relations (Eq.\ref{ucr1} for $\Delta p_0=0$), namely
\be
\Delta x \ge \frac{\hbar}{\Delta p} + \mbox{const} \cdot \Delta p
\label{ucrlit}
\ee
On the other hand this is probably the simplest modification
of the uncertainty
relation that implies nonzero minimal uncertainties in positions.
It is therefore quite likely to appear, at least as an
approximation, in any theory of
non-pointlike particles. Nevertheless, the coincidence of
these results may also indicate that our framework could be used
as an effective theory of an underlying theory
of explicitly non-pointlike particles like string theory.
\sn
More general than the above one-dimensional uncertainty relation
Eq.\ref{ucrlit} which is found in the literature, there appear
in our $n$- dimensional generalised commutation relations
Eqs.\ref{genHi}-\ref{genHf}
general matrices $B$ and $C$ which are related to
nonzero minimal
uncertainties in positions and further to a possible
noncommutativity of the position operators among themselves.
The introduction of such a matrices $B,C$ should be related to
effects at extremely short distances
like the Planck scale. We note that our general case is of the type
that is given in the literature for the one-dimensional case, but
we will here not speculate any further. In particular we can
not yet discuss implications of such ultraviolet modifications
on the external symmetry.

\subsection{Infrared generalisation of the uncertainty relations}
\label{grav}
In our generalised commutation relations
Eqs.\ref{genHi}-\ref{genHf} the matrices $A$ and $D$ have the potential
to introduce nonzero minimal uncertainties in momenta
and noncommutativity of the momentum operators among themselves.
We expected such large scale or 'infrared' effects to occur from
ordinary large scale gravity i.e. from curvature. We sketch
an ansatz that motivates the occurrence of a matrix $A$ from
large scale gravity:
\sn
Let us first explain why we intuitively expect
curvature to produce nonzero minimal
uncertainties in momenta or why it should at least modify
the uncertainty relations: Quantum theoretically, as one tries to
measure momentum more precisely, a particle's position gets
more uncertain. Momentum is a nonlocal property of a particle.
Measuring a momentum vector, in which vector space, i.e. in
the tangent space to which point of the manifold should it live?
Doing quantum mechanics on flat space,
one can identify all tangent spaces, however
this is no longer so in the presence of curvature due to the
path-dependence of parallel transport.
This should have an effect which becomes the more significant,
the more the wave function is spread over spacetime.
Intuitively we therefore expect this to yield a lower limit to the
precision of momentum determination. In order to make this precise,
we try an ansatz for how an observer $A$ at a point
$Q$ on a curved space could
define position and momentum operators, that would allow him to do
quantum theory in the vicinity of $Q$, including e.g. first order
curvature effects. Our strategy is, to
relate position- or momentum measurements
to vectors in the tangent- or cotangent space of $Q$.
\mn
The observer $A$ at a point $Q$ in
spacetime may set up a geodesic coordinate system $G_Q$ with
the event $Q$ at its
origin. $A$ can assign each event its 4-position coordinates with
respect to this frame. But how should $A$ define 4-momenta?
Classically as well as quantum theoretically,
momentum is the generator of translations, which means in practice
that it can be determined by measuring infinitesimal distances
(multiplied with the rest mass and divided by the infinitesimal
eigentime).
\sn
In curved space too, let $A$ define momentum as the generators of
translations. But how can $A$ measure (infinitesimal)
distances?
Just like in flat space, let $A$ define the distance
of two events E1 and E2 to be that 4-vector $\alpha\in T_Q$,
by which he has to
(infinitesimally) move from $Q$ to $Q^{\prime}$, so that when setting
up a parallel geodesic frame $G_{Q^{\prime}}$ there, the coordinates
of E1 in $Q1$ and E2 in $Q2$ coincide\footnote{Note that this yields
the definition of momentum as generating passive translations. Active
translations, 'moving all events by $\alpha$ within
 one coordinate frame'
would not be well defined
since the parallel transport of $\alpha$ to these events
would depend on the path of the parallel transport,
a fact which exactly
expresses the presence of curvature}.
\mn
Let us calculate the transformation of position coordinates
under the change to the geodesic coordinate
frame $G_{Q}$ from another 'infinitesimally by $\alpha$ translated'
geodesic frame $G_{Q^{\prime}}$, the axes of
which start off parallel to the axes of $G_Q$.
The origin $Q^{\prime}$ of the frame $G_{Q^{\prime}}$ has
 the coordinates
$\alpha^{\mu}$ in the frame $G_Q$. Trivially still
the coordinates ${x^{\prime}}^{\mu}$ of events
in the infinitesimal neighbourhood
of $Q^{\prime}$ (i.e. all
 ${x^{\prime}}^{\mu}{x^{\prime}}^{\nu}=0)$
transform into the frame $G_Q$
as $x^{\mu} =  \alpha^{\mu} + {x^{\prime}}^{\mu}$.
For events further away from the origin the transformation is
nontrivial since the (geodesic) axes of $G_Q$ and
 $G_{Q^{\prime}}$ are
then no longer parallel. Nevertheless all event's (passive)
coordinate transformations are well defined. We will here
calculate this transformation to the second order
${x^{\prime}}^{\mu}{x^{\prime}}^{\nu}$ in the distance
from the origin, which is of course the first order of length that
includes curvature.
\mn
Let us recall the construction of a geodesic frame or
of a 'Riemann normal coordinate system' around the point $Q$.
Using the exponential mapping each vector in the tangent space
of $Q$ is mapped to a point of the manifold
near\footnote{In general this mapping is unique as
far as the geodesics that start at $Q$ do not intersect.}
$Q$. The coordinates that we give this point
are just the coordinates of this tangent vector in an orthonormal
basis. Practically this means the following: Let us denote the
coordinates of a tangent vector by $x^{\mu}$. Which point in the
manifold has thus the coordinates $x^{\mu}$? It is the endpoint
of the geodesic which starts at $Q$ with the initial velocity
given by the tangent vector $x^{\mu}$
and evolved until the geodesic's parameter reaches 1.
\mn
Clearly in a geodesic frame the
geodesics through the origin are straight lines. E.g. the geodesic
which runs from the point $Q^{\prime}$ with an initial velocity
${x^{\prime}}^{\mu}$ to the event $E$ with the coordinates
${x^{\prime}}^{\mu}$
(where its geodesic parameter reaches 1) is a straight line
in the frame $G_{Q^{\prime}}$. We want to calculate $E$'s
coordinates in the frame $G_Q$ from its coordinates
${x^{\prime}}^{\mu}$ in the frame $G_{Q^{\prime}}$. To this end
we now simply calculate in the frame $G_Q$ the endpoint of this same
geodesic which starts at $Q^{\prime}$ with the initial velocity
${x^{\prime}}^{\mu}$, evolved until the geodesic's
 parameter reaches 1.
This is now nontrivial since our geodesic is not running through
the origin of $Q$ and is therefore not a straight line in the
frame $G_Q$. Taking
into account the first deviations from the straight line, which
occur at second order distance from $Q^{\prime}$ we obtain the
following result:
\sn
The coordinates $x^{\mu}$ in the frame $G_Q$ of
an event which has the coordinates ${x^{\prime}}^{\mu}$
in the frame $G_{Q^{\prime}}$ read to second order in $x^{\prime}$:
\be
x^{\mu} = \alpha^{\mu} + {x^{\prime}}^{\mu} - \frac{1}{2}\mbo
{\Gamma^{\mu}}_{\alpha\beta,\nu}(Q)\mbo \alpha^{\nu}
{x^{\prime}}^{\alpha} {x^{\prime}}^{\beta}
\label{trafo}
\ee
To see this, recall that $\Gamma(Q)=0$ in the frame $G_Q$
so that in this frame the geodesics through
the point $Q^{\prime}$ (which define the frame $G_{Q^{\prime}}$)
obey, to the given order, the equation:
\be
\frac{d^2x^{\mu}}{ds^2}+{\Gamma^{\mu}}_{\alpha\beta,\nu}(Q) \mbo
\alpha^{\nu} \frac{dx^{\alpha}}{ds}\frac{dx^{\beta}}{ds} = 0
\ee
So far our definition of momentum as generating infinitesimal
(passive) translations was classical. The observer $A$ at $Q$
can now also define the corresponding
quantum theoretical momentum generators $\p_{\mu}$
namely such, that they
generate just the translations given by Eq.\ref{trafo},
i.e. we require (recall that $\alpha$ is
infinitesimal i.e. $\alpha^2=0$)
\begin{eqnarray}
e^{-\frac{\alpha^{\nu}\p_{\nu}}{i\hbar}}\mbo \x^{\mu}\mbo
e^{\frac{\alpha^{\nu}\p_{\nu}}{i\hbar}}
  & = & \x^{\mu} + \frac{1}{i\hbar} \alpha^{\nu}
[\x^{\mu},\p_{\nu}]\\
  & = & \x^{\mu} + \alpha^{\mu} -
\frac{1}{2}\mbo
{\Gamma^{\mu}}_{\alpha\beta,\nu}(Q)\mbo \alpha^{\nu}
\x^{\alpha} \x^{\beta}
\label{exp2}
\end{eqnarray}
which means that the position and momentum operators of our
observer $A$ at the point $Q$ obey, to the first
gravity including order, the commutation relations:
\be
[\x^{\mu},\p_{\nu}] = i\hbar \delta_{\nu}^{\mu} - \frac{i\hbar}{4}
{\Gamma^{\mu}}_{\alpha\beta,\nu}(Q) \{ \x^{\alpha}, \x^{\beta} \}
\ee
Recall that the $\x^{\mu}$ are not the coordinate functions $x^{\mu}$,
which are not contravariant vectors. The newly defined
$\x^{\mu}$ and $\p_{\nu}$ are the components of covariant and
contravariant vectors at $Q$, which generate an abstract
Heisenberg algebra attached to $Q$ and which is to be represented
on a Hilbert space. We have to check that our definition is covariant.
\sn
Actually, at the origin of geodesic coordinate systems,
i.e. here at $Q$, there holds \cite{mtw}:
\be
{\Gamma^{\mu}}_{\alpha\beta,\nu}(Q) = -\frac{1}{3}
({R^{\mu}}_{\alpha\beta\nu} + {R^{\mu}}_{\beta\alpha\nu})
\ee
Using the antisymmetry of the Riemann curvature tensor $R$ in its
last two indices and its relation to the Jacobi curvature tensor $J$
\be
{J^{\mu}}_{\nu\alpha\beta} = \frac{1}{2} ({R^{\mu}}_{\alpha\nu\beta}
+ {R^{\mu}}_{\beta\nu\alpha})
\label{jr}
\ee
we find that the set of commutation relations is covariant
and has the simple form:
\be
[\x^{\mu},\p_{\nu}] = i\hbar \delta_{\nu}^{\mu} + \frac{i\hbar}{6}
{J^{\mu}}_{\nu\alpha\beta}(Q) \{ \x^{\alpha}, \x^{\beta} \}
\label{dh}
\ee
Note that the Riemann curvature tensor can also be
expresed in terms of the Jacobi curvature tensor
${R^{\mu}}_{\alpha\nu\beta} = 2/3\mbo ({J^{\mu}}_{\nu\alpha\beta}
-{J^{\mu}}_{\beta\alpha\nu})$
which means together with Eq.\ref{jr}
that the Jacobi curvature tensor
carries exactly the same information content
as the Riemann curvature tensor.
\sn
We are therefore lead to suggest the following:
\sn
An observer $A$ located at some arbitrary point $Q$ in
curved spacetime uses the Heisenberg algebra with
commutation relations as defined at $Q$ by the Eq.\ref{dh}.
It should allow him to do quantum mechanics and quantum field theory
in the vicinity of $Q$, namely up to the
second order\footnote{Continuing the expansion in
Eqs.\ref{trafo},\ref{exp2}
yields higher powers in the $\x$'s on the rhs of Eq.\ref{dh}
which expand the validity of Eq.\ref{dh} to higher orders
in the distance $x^{\mu}$ from $Q$}
in the distance $x^{\mu}$ from $Q$.
This is a nontrivial ansatz for quantum theoretical
momentum on curved space
and we will have to compare it to other approaches
where momentum is defined e.g. via (ordinary)
Fourier transformation in an approximation or e.g. via
the vanishing covariant derivative of some stress energy
tensor, see e.g. \cite{bd,LesHouches}.
Let us note that our definition of momentum as generating
(passive) translations obeys at least the
following three criteria:
It is well defined how it is practically to be measured, it has
the right quantum theoretical limit for vanishing curvature,
and it also has the right classical limit for $\hbar\rightarrow 0$,
where it is still the generator of
passive translations on classical curved space, which was
our definition of momentum there.
\sn
We gave an ansatz for how curvature, in
an approximation, could kinematically
influence the Heisenberg commutation relations of position and
momentum operators. Here we will not
yet study the dynamics of example systems nor will we
try to express how stress-energy expectation values
determine the curvature $J$.
Work in this direction is in progress.
Let us however discuss what it would mean
for the uncertainty relations
if our definition of momentum proves to be applicable:
\mn
The above commutation relations Eq.\ref{dh} are of the form
of our generalised Heisenberg algebras
 Eqs.\ref{genHi}-\ref{genHf},
with a nonvanishing matrix $A$, being the Jacobi tensor.
However, from ${J^{\mu}}_{(\alpha\beta\gamma)}=0$
it is clear that these commutation relations are
 not precisely of the form
of the examples of commutation relation for which we
had Hilbert space representations.
For field theoretical studies, we will therefore
 first have to construct
Hilbert space representations for our commutation
relations Eq.\ref{dh},
which are of course not necessarily
Bargmann Fock representations.
Can we already say something about the occurrence of
nonzero minimal uncertainties in momenta? Note that
because of ${J^{\mu}}_{\mu\mu\mu}= 0$,
nonzero minimal uncertainties in momenta will not directly appear
from an equation of the type of Eq.\ref{ucr1}. However, there
are in general nonvanishing elements ${J^{\mu}}_{\nu\mu\mu}$ which
have the potential to produce nonzero minimal uncertainties
in the momentum $\p_{\nu}$ through uncertainty relations of the type
$\Delta x^{\mu}\Delta p_{\nu} \ge \frac{\hbar}{12}
{J^{\mu}}_{\nu\mu\mu} (\Delta x^{\mu})^2 + \mbox{rest}$.
A detailed analysis is in progress.
\mn
Let us discuss implications on the external symmetry that could
be expected from our ansatz. Our momentum operators would provide
the Poincar{\'e} generators of (passive)
translations and also
the Poincar{\'e}
generators of (passive) rotations should be
contained in the
Heisenberg algebra, i.e. it should be possible to express them
in terms of the position and momentum operators.
We had already mentioned that nonzero minimal
uncertainties, in particular in energy,
would e.g. make a strict distinction
between positive and negative frequencies impossible, so that the
propagators could reflect spontaneous particle creation
and curvature induced energy (and momentum) nonconservation.
\sn
Of course, energy and momentum are not conserved on curved
space since gravity is described as an intrinsic
property of spacetime itself. The covariant
divergence of the stress energy tensor vanishes, however, the
covariant divergence decomposes
into the ordinary divergence plus a term that involves the field
strength $\Gamma$.
\sn
Related to this, it is on a general curved spacetime
not possible to define a total momentum vector
or e.g. a total angular momentum tensor for a
collection of (even classical) particles.
How would this fact be reflected
in our framework? It is that the additive coalgebra structure is
no longer
compatible with the algebra structure of the generalised
Heisenberg algebra. Let us explain this important point.

For each particle separately one can consider its position
or momentum. Practically this means that the information on its
position and momentum is encoded in a vector which is contained
in the representation of a Heisenberg algebra of
position and momentum
operators. Each particle has its Heisenberg algebra acting on
its states. Considering $n$ particles,
or more precisely, considering $n$- point functions,
there are $n$ Heisenberg
algebras acting on the tensor product of their
 $n$ representation spaces.
So far this trivially expresses that one can
speak of each particle's,
or leg's, momentum, position etc separately.
How is it expressed when there is also e.g. a
total momentum? If the
collection of the particles can be considered
as a new 'total' system,
it mathematically means that there is \it one
 \rm Heisenberg
algebra which not only acts on each particle
representation separately but
also on their tensor product representation space.
E.g. the operator $\p_{\mu}$ that acts on one
representation must therefore
also act on the tensor product of two
 representations, namely as
$\Delta \p_{\mu} = \p_{\mu} \otimes 1 +
1 \otimes \p_{\mu}$, so that
the two momenta are added. The
comultiplication\footnote{A comultiplication
 obeys the axioms dual
to the axioms of an algebra i.e. with the
direction of the arrows reversed,
thus being coassociative, having a counit
etc., see e.g. \cite{sweedler,abe}}
$\Delta$ that
maps the Heisenberg algebra $H$ onto its
 tensor square
$\Delta: H \rightarrow H \otimes H$ can be iterated to map onto
arbitrary high tensor powers of $H$, so that the
 Heisenberg algebra can
act on arbitrary many legs to determine total momentum, total
angular momentum etc.. For this to be consistent, the
comultiplication
has to be an algebra homomorphism, i.e. it should not matter whether
one first applies the comultiplication and then uses the commutation
relations or vice versa. While this is the case for the
ordinary Heisenberg algebra on flat space,
expressing that e.g. total
momentum is well defined, it is not difficult to see, that
the additive comultiplication is no
longer consistent with our generalised
commutation relations Eqs.\ref{dh}. This has of course to be expected
since on curved space collections of particles can no longer be
described as one particle with 'total' properties.
\sn
Note that even in the special case when the commutation relations
are preserved by a quantum group (which itself has a consistent
coproduct structure), the Heisenberg algebra is not consistent
(i.e. is not a bialgebra) with the coproduct that would describe
additivity.
Recall also that our quantum mechanical
wave functions $\phi_r$
as well as our fields $\phi_{\vec{r}}$ that we summed over
in the path integral remain strictly commutative.
The Bargmann Fock
functions $\phi(\etab)$ are functions of operators
and their noncommutativity is merely the noncommutativity
of the generalised Heisenberg algebra and does not imply
noncommutativity of the wave functions. This
means in particular that, unlike other ansatzes, we do
not describe particles with braid statistics, not
even with the sub class of Heisenberg commutation relations
that are preserved under the action of a quantum group.
\sn
The connection between spin and statistics on a general curved, and
in the ultraviolet possibly noncommutative space must be intimately
related to unitarity and the action of the Poincar{\'e}
generators. We will here not further discuss these questions.
The aim of this section was merely to sketch a possible
ansatz for generalised Poincar{\'e} operators, namely as generators
of passive translations and rotations, and to show that
this naturally leads to the type of generalised commutation relations
that we had considered in the previous sections.

\section{Summary and Outlook}

We followed the idea that gravity puts (dynamical)
lower limits to the
precision to which positions or momenta can be determined, and
studied whether
this could lead to a natural regularisation of divergencies.
Technically we made the ansatz that these
lower limits of precision take the
form of nonzero minimal
uncertainties, where we applied the usual quantum theoretical
definition of the uncertainty in an observable.
\sn
The ordinary quantum field
theoretical framework does not have lower limits
to the accuracy to which positions or momenta could
formally be described. This feature is
independent of the actual
interactions since uncertainty relations are a purely
'kinematical' consequence of the Heisenberg commutation rules.
\sn
Therefore, in order to introduce nonzero minimal uncertainties
in positions and momenta, we defined generalisations of
the quantum theoretical framework itself, by generalising the
Heisenberg algebra of positions and momenta.
After developing
useful algebraic techniques in the quantum mechanical case, we
proceeded to the study of quantum field theory. It was crucial
to realise that in quantum field theory too the Heisenberg
algebra of 4-positions and 4-momenta 'sets the stage':
Points e.g. in position space are points in the continuous
spectra of the position operators and e.g. the scalar
product of position and momentum eigenvectors defines the
unitary transformation from position to momentum space.
A field is an abstract vector in the
representation space of the Heisenberg algebra.
\sn
Choosing the Bargmann Fock basis for the
fields\footnote{Recall that while it is usual practice to choose the
Bargmann Fock basis in the space in which the fields take their
values, we are here choosing the Bargmnn Fock basis in the space
on which the fields are defined.} i.e. for the vectors
in this representation space,
we reformulated the ordinary quantum field theoretical
path integral in this new basis.
No longer needing or assuming a position or a momentum
eigenbasis, we were then able to
generalise the Heisenberg algebra. In the general case
the position operators (as well as the momentum operators)
no longer commute even among themselves, i.e. 'position
and momentum space become noncommutative'.
As a consequence 'points' in position or momentum
space i.e. position or momentum eigenvectors are abandoned and
instead there naturally appear nonzero minimal uncertainties
in positions and momenta.
\sn
We were led to
a large class of generalisations of the Heisenberg algebra
which imply these nonzero minimal uncertainties.
Representations for two
sub classes of these generalised Heisenberg algebras were
supplied by algebraic techniques that have been
developed in the field of quantum groups. We could
therefor in two cases
work out explicitly the Feynman rules and the example of a
loop integration in scalar $\phi^4$ theory.
In both cases we showed the regularisation of a
quadratically ultraviolet divergent graph through
the generalisation.
\sn
This supports the idea that the introduction of nonzero minimal
uncertainties provides a new regularisation method. We do not
yet have a general proof. Given representations
not only of our two sub classes but also for
the general Heisenberg algebras described
by Eqs.\ref{genHi}-\ref{genHf}, it
should be straightforward to work out also the
 correspondingly generalised
quantum field theoretical framework and to check the convergence
of loop integrations.
\sn
Thereby the corresponding generalisations
of the framework itself are determined by
sticking to the definition
that the path integral is the sum over all 'fields', i.e. the sum
over all vectors in the representation of the Heisenberg algebra
of 4-positions and 4-momenta. It is of course not
necessary that the Heisenberg commutation relations
are conserved by the action of a quantum group, as it is the
case in the first of the two sub classes of Heisenberg commutation
relations that we worked with.
\sn
Once representations of our generalised Heisenberg
algebras Eqs.\ref{genHi}-\ref{genHf} are found,
our framework offers many possibilities to formulate
and to study completely new ansatzes for locality in the
ultraviolet. As we said, one can possibly even extend
the beautyful position and
momentum duality from the kinematical to the dynamical level,
which would relate the very small to the very large scales
in a natural way.
\sn
Another important issue for further study is the fact
 that due to ordering ambiguities it is not
straightforward to find the 'correct' generalisation of the
action functional.
A guiding principle for the generalisation of Lagrangians
must be found on physical grounds
(one may even speculate that such a principle could single out
Lagrangians that have
generalisations i.e. ultimately, which could be consistent with
the description of gravity). In particular, through
the introduction of nonzero minimal uncertainties
in positions, the notions of 'pointlike'
particles, of point-interaction and of locality will be generalised.
In our calculations we have so far only used a simple
working hypothesis for the generalised locality.
\sn
We should keep in mind that among the main criteria for the exploration
of the new possibilities must be the conservation of the
basic features of quantum field theory
microcausality and unitarity. Also the relation
between spin and statistics has to be clarified.
Like in ordinary quantum field theory it is therefor
crucial to understand the r{\^o}le of the (local) Poincar{\'e} algebra of
translation and rotation generators. It is contained
in the Heisenberg algebra and we must therefor
study first the origin of the generalised commutation relations.
\sn
To this end we discussed possible anatzes to make the
origins of ultraviolet and infrared
modifications of the Heisenberg
algebra explicit. For the modifications that imply nonzero minimal
uncertainties in positions we found motivation in the
literature, coming from studies in quantum gravity and string
theory where uncertainty relations of exactly our type had
already appeared. We also proposed an ansatz for
how large scale gravity enters into the definition of
the momenta as the generators of passive translations.
One then indeed straightforwardly arrives at infrared
modified commutation relations of the expected form, which have
the potential to produce dynamical nonzero minimal uncertainties in
momenta. We explained our further programme in this direction,
namely to study
the local Poincar{\'e} algebra generated by momenta and angular
momenta etc. which are identified as
the generators of passive translations and rotations.
\mn
As we said, our framework could be considered as being simply
an approach towards a new regularisation method.
However, we are here not the first to suggest that
new physical insight could be gained by studying the interplay of
quantum theory, the measurement process and gravity,
see e.g.\cite{haw-pen}. Although quite different from
previous ansatzes, also our ansatz of the use of
noncommutative geometry with its nonzero minimal uncertainties
can probably be interpreted in this way.
\sn
Ultimately we would expect the fully developed formalism to express a
dynamical interdependence of gravity and quantum
theory: On the one hand the 'local'
curvature would influence the 'local' Heisenberg commutation
relations. On the other hand the generalised Heisenberg algebra
would influence the very concept of locality by producing
modified uncertainty relations and
nonzero minimal uncertainties in positions and momenta.
Both, ordinary quantum theoretical
commutation relations as well as ordinary curved manifolds would
only exist as limiting cases and
the framework could therefore probably be called
being of type IV in the discussion given in \cite{isham}.
\sn
We can here only offer preliminary results, however,
any possibility that gravity effects could, instead of
rendering quantum field theories nonrenormalisable, actually
make them finite, should be worth exploring.
\mn
\bf Acknowledgement: \rm It is a pleasure to thank many members of
our department for the stimulating atmosphere,
for their interest and for their criticisms, in
particular T. Brzezi\'nski, S.B. Davis,
G. Gibbons, M.B. Green, A. Kent,
P.V. Landshoff, A. Macfarlane,
S. Majid, M. Perry, T. Samols, A.R. Steif, J.C. Taylor and G.M.T. Watts.

\end{document}